\documentclass[12pt]{article}
\usepackage{amssymb}
\usepackage{epsf}
\usepackage{epsfig}
\usepackage{psfrag}
\usepackage{rotating}
\newcommand{\lsim}{
\mathrel{\hbox{\rlap{\hbox{\lower4pt\hbox{$\sim$}}}\hbox{$<$}}}}

\newcommand{\gsim}{
\mathrel{\hbox{\rlap{\hbox{\lower4pt\hbox{$\sim$}}}\hbox{$>$}}}}

\newcommand{\vtd}{|V_{td}|}

\newcommand{\be}{\begin{equation}}
\newcommand{\ee}{\end{equation}}
\newcommand{\bi}{\begin{itemize}}
\newcommand{\ei}{\end{itemize}}
\newcommand{\ord}{{\cal O}}

\newcommand{\RE}{{\rm Re}}
\newcommand{\IM}{{\rm Im}}

\def\klpn{K_{\rm L}\rightarrow\pi^0\nu\bar\nu}

\begin{document}
\begin{titlepage}
\vspace*{-0.5truecm}

\begin{flushright}
TUM-HEP-562/04\\
hep-ph/0410309
\end{flushright}

\vspace*{0.3truecm}

\begin{center}
\boldmath

{\Large{\bf Particle-Antiparticle Mixing, $\varepsilon_K$ and 

\vspace{0.3truecm}

the Unitarity Triangle in the Littlest Higgs Model}}
\unboldmath
\end{center}

\vspace{0.4truecm}

\begin{center}
{\large\bf Andrzej J. Buras, Anton Poschenrieder
and Selma Uhlig} 
\vspace{0.4truecm}

{\sl Physik Department, Technische Universit\"at M\"unchen,
D-85748 Garching, Germany}

\vspace{0.2truecm}

\end{center}

\vspace{0.4cm}
\begin{abstract}
\vspace{0.2cm}\noindent
{\small We calculate 
the $K^{0}-\bar{K}^{0}$, $B_{d,s}^{0}-\bar{B}_{d,s}^{0}$} 
mixing mass differences $\Delta M_K$, $\Delta M_{d,s}$ and the 
CP-violating parameter $\varepsilon_{K}$ in the Littlest Higgs (LH) model. 
For $f/v$ as low as $5$ and the Yukawa parameter $x_L<0.8$, the enhancement of 
$\Delta M_{d}$ 
amounts to at most $20\%$. Similar comments apply to $\Delta M_s$
and $\varepsilon_{K}$.
The correction to $\Delta M_{K}$ is negligible.
 The dominant new contribution in this parameter range, 
calculated here for the first time, 
comes from the box diagrams with ($W_L^\pm,W_H^\pm$) exchanges and 
ordinary quarks that are only suppressed by the mass of $W_H^\pm$ 
but do not involve explicit $\mathcal{O}(v^2/f^2)$ factors. 
This contribution is strictly positive.
The explicit $\mathcal{O}(v^2/f^2)$ corrections to the SM diagrams
with ordinary quarks and two $W_L^\pm$ exchanges have to be combined with 
the box diagrams with a single heavy $T$ quark exchange 
for the GIM mechanism to work. 
These $\mathcal{O}(v^2/f^2)$ corrections are found to be of 
the same order of magnitude as the ($W_L^\pm,W_H^\pm$) contribution but 
only for $x_L$ approaching $0.8$ they can compete with it. 
We point out that for $x_L>0.85$ box diagrams with two $T$ exchanges have to
be included. Although formally $\mathcal{O}(v^4/f^4)$, 
this contribution is dominant for $x_L \approx 1$ due to non-decoupling of $T$
that becomes fully effective only at this order.
We emphasize, that the concept of the unitarity triangle is still useful
in the LH model, in spite of the $\mathcal{O}(v^2/f^2)$ corrections
to the CKM unitarity involving only ordinary quarks. We demonstrate 
the cancellation of the divergences in box diagrams that appear when one 
uses the unitary gauge for $W_L^\pm$ and $W_H^\pm$. 

\end{abstract}

%
%
%
\end{titlepage}

\thispagestyle{empty}
\vbox{}

\setcounter{page}{1}
\pagenumbering{roman}



\setcounter{page}{1}
\pagenumbering{arabic}
\section{Introduction}\label{sec:intro}
An attractive idea to solve the gauge hierarchy problem is to regard 
the electroweak Higgs boson as a pseudo-goldstone boson of a certain global 
symmetry that is broken spontaneously at a scale 
$\Lambda \sim 4 \pi f \sim \mathcal{O}\left(\textrm{10 TeV}\right)$, 
 much higher than the vacuum expectation value $v$ of the standard Higgs 
doublet. Concrete realizations of this idea are the ``Little Higgs'' 
models \cite{LH1}-\cite{LH5} in which the Higgs field remains light, 
being protected by 
the approximate global symmetry from acquiring quadratically divergent 
contributions to its mass at the one-loop level. In models of this type 
new heavy particles are present, that analogously to supersymmetric 
particles allow to cancel the quadratic divergences in question.
Reviews of the Little Higgs models can be found in \cite{LHREV}.

One of the simplest models of this type is the ``Littlest Higgs'' model 
\cite{LH4} (LH) in which, in addition to the Standard Model (SM) particles, 
new charged heavy vector bosons ($W_H^\pm$), a neutral heavy vector 
boson ($Z_H$), a heavy photon ($A_H$), a heavy top quark ($T$) and a 
triplet of heavy Higgs scalars ($\Phi^{++}$, $\Phi^{+}$, $\Phi^{0}$) 
are present. The details of this model including the Feynman rules have been 
worked out in \cite{Logan} and the constraints from various processes, 
in particular from electroweak precision observables and direct new particles
searches, have been extensively discussed in \cite{Logan}-\cite{PHEN6}. 
It has been found that 
except for the heavy photon $A_H$, that could still be as ``light'' as 
$500~\textrm{GeV}$, the masses of the remaining particles are constrained 
to be significantly larger than $1~ \textrm{TeV}$.

The question then arises whether the Flavour Changing Neutral Current
(FCNC) processes, such as particle-antiparticle mixings and various
rare $K$ and $B$ decays, that played such an essential role in the
construction of the SM, could further constrain the parameters
of the LH model. This issue is particularly interesting because the
mixing of the SM top quark ($t$) and of the heavier top ($T$) induces
violation of the three generation unitarity of the CKM matrix at 
$\ord(v^2/f^2)$ that
is essential for a natural suppression of the FCNC processes (GIM
mechanism) \cite{GIM}.
Moreover, as the mass of $T$ must be larger than $1~ \textrm{TeV}$, 
interesting non-decoupling effects of this very heavy quark could play a role
similar to the non-decoupling of $t$ from FCNC processes that increases 
quadratically with $m_t$.

In the present paper we calculate the new particle contributions to
$K^0-\bar K^0$, $B^0_{d,s}-\bar B^0_{d,s}$ mixings and to the CP
violation parameter $\varepsilon_K$ within the LH model \cite{LH4}. We also
address the unitarity triangle in the presence of the violation of the
CKM unitarity at the $\ord(v^2/f^2)$ level, pointing out that this triangle 
can be used in the LH model, provided the uncorrected CKM elements are used 
as basic parameters. The corresponding analysis
of rare $K$ and $B$ decays, that is more involved, will be presented
elsewhere \cite{DEC05,BPU05}.

We are not the first to address the question of FCNC processes within
the LH model. 
In \cite{BSG} the LH corrections to the the decay $B\to X_s\gamma$ have 
 been found to be small, while in \cite{QUAD} it has been pointed out
that sizable effects could be present in $D^0-\bar D^0$ mixing, where 
in contrast to  processes involving external down quarks, FCNC transitions 
are already present at the tree level.
Recently, in two interesting papers, Choudhury et
al. \cite{IND1,IND2} analyzed the  $B^0_{d}-\bar B^0_{d}$ mass difference 
$\Delta M_d$ and the decay $\klpn$ within the model in question, finding a
significant suppression of $\Delta M_d$ and a large enhancement of the
branching ratio for $\klpn$ relative to the SM expectations.

Unfortunately our analysis of $\Delta M_d$ presented here does not
confirm the findings of \cite{IND1}, both in sign and magnitude, for the
same input parameters. Instead of a  suppression of $\Delta M_d$ found 
by these authors, we
find an {\it enhancement} in the full range of parameters considered
but this enhancement amounts to at most $20 \%$ for $f/v \ge 5$,
the masses of $W_H^{\pm}$, $T$ and $\Phi^{\pm}$ larger than $1.5~{\rm  TeV}$
and the Yukawa parameter $x_L<0.8$ (see (\ref{XL})).
The same comments apply to 
$\Delta M_s$ and $\varepsilon_K$. The corrections to 
$\Delta M_K$ are negligible.
We conclude therefore that in view of non-perturbative
uncertainties in $\Delta M_K$, $\Delta M_{d,s}$ and $\varepsilon_K$ 
it will be very difficult in this range of parameters 
to distinguish the LH expectations for these quantities
 from the SM ones. Consequently, in contrast to \cite{IND1}, we find 
that the constraints on LH model parameters coming from $\Delta M_d$ 
are for $x_L<0.8$ substantially weaker than the ones coming from 
other processes 
\cite{Logan}-\cite{PHEN6}.  
On the other hand, as pointed out in \cite{DEC05} and below, for $x_L>0.85$, 
where the non-decoupling effects of $T$ enter at full strength, the LH 
corrections turn out to be larger, putting some constraints on the space of 
parameters \cite{DEC05}.

The first difference between our analysis and the one of \cite{IND1} is 
that we include
the box diagrams with the ordinary quarks, one 
$W_L^{\pm}$ and one $W_H^{\pm}$ exchanges that enter $\Delta M_d$ 
at $\ord(1)$ in
the couplings and are only suppressed by the mass of $W_H^{\pm}$
relatively to the usual box diagrams with two $W_L^{\pm}$
exchanges. Surprisingly the authors of \cite{IND1} omitted this
contribution although they took into account partially the
$\ord(v^2/f^2)$ corrections to the box diagrams with
($W_L^{\pm}$,$W_H^{\pm}$) exchanges. While we find the latter
contribution totally negligible, the former turns out to be the
dominant one for $x_L<0.7$ and, being positive, governs the sign of 
the full effect.

The second important contribution, also considered in \cite{IND1}, are the
$\ord(v^2/f^2)$ effects related to the modification of the vertices in
the usual box diagrams with ordinary quarks and two
$W_L^{\pm}$ exchanges. As emphasized in \cite{IND1,IND2}, 
due to the 
violation of the
CKM unitarity at $\ord(v^2/f^2)$ these corrections have to be
considered simultaneously with box diagrams involving single $T$ for
the GIM mechanism to be effective. We find that these $\ord(v^2/f^2)$
contributions can have both signs depending on the input parameters
but in a large region of parameters considered they interfere
constructively with the diagrams with ($W_L^{\pm}$,$W_H^{\pm}$)
exchanges, increasing the enhancement of $\Delta M_d$, $\Delta M_s$
and $\varepsilon_K$. The general structure of this contribution presented,
 before the use of the GIM mechanism, in \cite{IND1} is equal to ours but 
their final numerical result indicates that the sign of this contribution 
and also its magnitude differ from our findings.

The third contribution, not considered in \cite{IND1},
comes from box diagrams with two 
$T$ exchanges. Although formally $\ord(v^4/f^4)$  
this contribution 
increases linearly with
$x_T=m_T^2/M^2_W$ and with $x_T=\ord(f^2/v^2)$  constitutes effectively
an $\ord(v^2/f^2)$ correction. 
For the Yukawa coupling parameter $x_L\approx 1$, this contribution 
turns out to be  more important than the remaining
$\ord(v^2/f^2)$ corrections. In particular it is larger than 
the $\ord(v^2/f^2)$ 
contribution of box diagrams with a single 
$T$ exchange discussed above that increases only logarithmically with
$x_T$.
 
We are aware of the fact that with increasing $m_{T}$ also one-loop 
corrections to the SM Higgs mass increase. 
Typically for $m_{T} \geq 6~ \textrm{TeV}$ a fine-tuning of at least 
$1\%$ has to be made in order to keep $m_{H}$ below $200\,\textrm{GeV}$
\cite{Csaki:2003si,Perelstein:2003wd}. 
As roughly $f/v \geq 8$ is required by electroweak precision studies 
\cite{Logan}-\cite{PHEN6},
the non-decoupling effects of $T$ considered here can be significant 
and simultaneously consistent with 
these constraints only in a narrow range of $f/v$. 
But these bounds are clearly  model 
dependent \cite{Chang:2003un,Chang:2003zn} and  we will consider the range 
$5 \leq f/v \leq 15$ and $x_L\le 0.95$ for completeness.

The fourth non-negligible correction, not considered in \cite{IND1}, is the one
related to the use of the standard value of the Fermi constant $G_F$ that
enters quadratically in all the quantities considered here. In order to 
include this correction in the 
evaluation of $\Delta M_i$ and $\varepsilon_K$, we calculate 
the amplitude for the muon decay in the LH model at the tree level. The 
resulting additional correction to $\Delta M_{d,s}$, $\varepsilon_K$, and
$\Delta M_K$ amounts to at most a few percent  
but being negative it reduces the enhancements slightly. 

Finally, we find that the contribution of the heavy scalar
$\Phi^{\pm}$ can be neglected for all practical purposes as it is well below
$1 \%$ of the full result for all quantities considered.

On the technical side, we have performed the calculations in the
unitary gauge for the $W_L^{\pm}$ and $W_H^{\pm}$ propagators which
has the nice virtue that only exchanges of physical particles have to be
considered. On the other hand in contrast to a $R_{\xi}$ gauge with a finite
gauge parameter $\xi$, the box diagrams in the unitary gauge are
divergent, both in the SM and the LH model. As already stated in
\cite{IND1} these divergences cancel when the unitarity of the CKM matrix
in the SM is used and the contribution of the heavy $T$ is included in
the LH model at
$\ord(v^2/f^2)$. As the authors of \cite{IND1} did not demonstrate this
explicitly, we will show this cancellation in Section \ref{sec:GIM}.
This exercise turned out to be very instructive. Indeed, the
cancellation of the divergences in box diagrams at  $\ord(v^2/f^2)$
takes only place when the vertex involving $W_L^{\pm}(W_H^{\pm})$
and $\bar T d_i$ with $d_i$ being ordinary down quarks, has the same
factor $i$ as the vertex involving the weak gauge bosons and $\bar t d_i$. 
This is not fully evident from  the widely used Feynman rules for
the LH model given in \cite{Logan} that uses different phase conventions for 
the $T$ and $t$ fields.

Our paper is organized as follows. In 
Section \ref{sec:LH} we recall very briefly those elements of the LH model
that are necessary for the discussion of our calculation. As a preparation 
for subsequent sections we calculate the amplitude for the muon decay in 
the LH model at the tree level and we investigate whether the usual
determination of the CKM elements, not involving the top quark, by means of 
tree level decays could be affected by the LH contributions in a
non-negligible manner. This turns out not to be the case. 
Performing analogous exercise for the tree level decay of the top quark into 
$b$ quark and leptons, we demonstrate how in principle the violation of the 
three generation CKM unitarity in the LH model could be detected
experimentally. Finally we emphasize that working with uncorrected CKM
elements as basic parameters, allows to display the effects of the LH
 contributions in the usual $(\bar\varrho,\bar\eta)$ plane \cite{WO,BLO}. 
They manifest
themselves primarily in the modification of the angle $\gamma$ and the side
$R_t$ in such a manner that the angle $\beta$ and the side $R_b$ remain
unchanged.  While this analysis is partly academic in view of the smallness
of corrections found here, it could turn out to be useful in other processes 
and other Little Higgs models in which larger effects in FCNC processes 
could be  present.  

In Section \ref{sec:GIM} we demonstrate explicitly the
cancellation of the divergences in the box diagrams calculated in the
unitary gauge. In Section \ref{sec:AR} we
discuss briefly our calculation for $x_L\le 0.8$ and present analytic 
expressions for
the relevant  contributions in this parameter region. For completeness we give
in Appendix A the results for the $\ord(v^2/f^2)$ corrections to box diagrams 
with $(W_L^\pm,W_H^\pm)$ exchanges and 
the
contribution of the scalars $\Phi^{\pm}$. It will be clear from these
formulae that these corrections are fully negligible. 
In Section \ref{DEC} we discuss the non-decoupling effects of $T$, that are
already visible in the box diagrams with a single $T$ exchange considered 
in Section \ref{sec:AR}, but are fully effective only in the region $x_L\approx 1$
in which the dominant correction $\ord(v^4/f^4)$, the box diagram with two 
$T$ exchanges, has to be taken into account.

In Section \ref{sec:numerics}
we present the numerical analysis of the formulae of Sections \ref{sec:AR}
and \ref{DEC}.
In Section \ref{SEC6}
 we briefly discuss the issue
of QCD corrections within the LH model. For scales $\mu\le \mu_t=\ord(m_t)$ 
they are 
the same as in the SM but the contribution of QCD corrections from higher 
scales are different.
In view of the smallness of
the new contributions and the theoretical uncertainties involved, it
is clearly premature to compute these additional QCD corrections. Still our
discussion indicates that they should further
suppress the LH contributions. A brief
summary of our paper is given in Section \ref{SEC7}.

\section{Aspects of the Littlest Higgs Model}\label{sec:LH}
\setcounter{equation}{0}
\subsection{Preliminaries}
Let us first recall certain aspects of the LH model that are relevant for our
work. The full exposition can be found in the original paper \cite{LH4} and 
in \cite{Logan}, 
where Feynman rules for the LH model have been worked out. We will follow the
notations of \cite{Logan}, although due to different phase conventions for the $t$
and $T$ fields, our rules for the vertices $W_L^\pm\bar T d_j$ and 
$W_H^\pm\bar T d_j$ differ by a crucial factor $i$ as discussed below.
 
The new particles that enter the calculations in the present paper are 
$W^\pm_H$, $T$ and $\Phi^\pm$. To the order in $v/f$ considered, 
their 
masses and their interactions with ordinary quarks and leptons can be 
entirely expressed in terms of
\be\label{ninput}
m_{t}\equiv \overline{m}_t(m_t) = 
168.1~\textrm{GeV}, \quad M_{W^\pm_{L}} = 80.4~ \textrm{GeV},
\quad M_H\ge 115~\rm{GeV}
\ee
and the following three new parameters of the LH model
\be
{f}/{v}, \quad s, \quad x_{L}.
\ee
Using the formulae in \cite{Logan} we find
\be\label{mass}
m_{T} =  \frac{f}{v} \frac{m_t}{\sqrt{x_{L}\left(1 - x_{L}\right)}}, 
\quad  M_{W^\pm_{H}}= \frac{f}{v} \frac{M_{W^\pm_L}}{s c}, \quad
M_{\Phi^\pm}\ge \sqrt{2} M_H\frac{f}{v}.
\ee
As $\Phi^\pm$ will play a negligible role in this analysis, we need only to
know its lower bound.

\begin{figure}[h]
\vspace{0.10in}
\epsfysize=2.0in
\centerline{
\epsffile{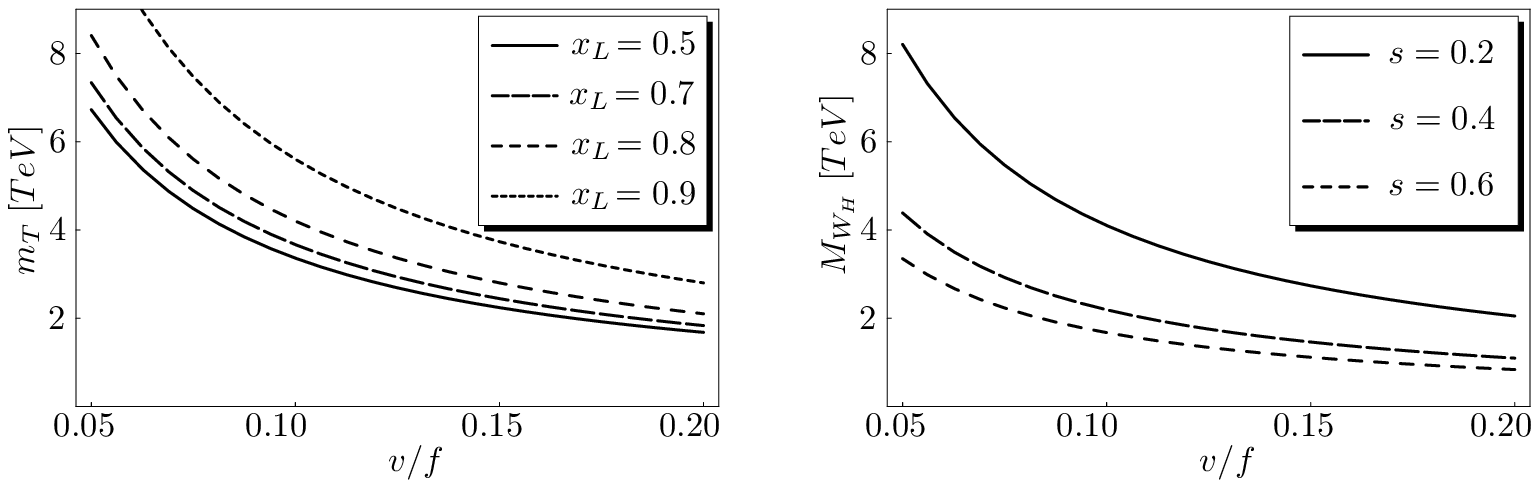}}
\vspace{0.08in}
\caption{The masses of the heavy top quark $T$ and the heavy $W_H$ boson
  as functions of $v/f$ for different values of $x_L$ and $s$.}
\label{fig:masses}
\end{figure}

We recall that $v=246~\rm{GeV}$ is the vacuum expectation value of the
standard Higgs doublet. The parameters $s$ and $c$ are the sine and the
cosine  of the mixing
angle between $SU(2)_1$ and $SU(2)_2$ gauge bosons of the original gauge
symmetry group $[SU(2)_1\otimes U(1)_1]\otimes[SU(2)_2\otimes U(1)_2]$
that is spontaneously broken down to the SM gauge group. The resulting
$SU(2)_L$ gauge coupling $g$ is then related to the $g_i$ couplings of the 
$SU(2)_i$ groups through
\be
g= s g_1=c g_2, \qquad  c^2=1-s^2.
\ee
Finally, 
\be\label{XL}
x_L=\frac{\lambda_1^2}{\lambda_1^2+\lambda_2^2},
\ee
where $\lambda_1$ is the Yukawa coupling in the $(t,T)$ sector and 
$\lambda_2$ parametrizes the mass term of $T$. The parameter 
$x_{L}$ enters the sine of the $t$-$T$ mixing which is  simply given 
by $x_{L} v/f$. As we will see below, the
parameter $x_L$ describes together with $v/f$ the size of the violation of
the three generation CKM unitarity and is also crucial for the gauge
interactions of the heavy $T$ quark with the ordinary down quarks. 
$\lambda_i$ are expected to be $\ord(1)$ with \cite{Logan} 
\be
\lambda_i\ge \frac{m_t}{v}\,, \qquad {\rm or}\qquad
\frac{1}{\lambda_1^2}+ \frac{1}{\lambda_2^2}\approx
\left(\frac{v}{m_t}\right)^2
\ee
so that within a good approximation
\be\label{lambdarel}
\lambda_1=\frac{m_t}{v}\frac{1}{\sqrt{1-x_L}}\,,\qquad
\lambda_2=\frac{m_t}{v}\frac{1}{\sqrt{x_L}}\,.
\ee
$x_L$ can in principle vary in the range $0<x_L<1$. For $x_L\approx 0$ and
$x_L\approx 1$, the mass $m_T$ becomes very large.
This is seen in Fig.~\ref{fig:masses}, where we show 
the masses of the heavy top quark $T$ and the heavy $W_H$ boson
  as functions of $v/f$ for different values of $x_L$ and $s$. The fact 
that for $x_L \approx 1$ the Yukawa coupling $\lambda_1$ becomes large 
is responsible for the non-decoupling of $T$ at fixed $v/f$ as discussed 
in Section \ref{DEC}.

\subsection{Fermion-Gauge Boson Interactions}
In table VIII of \cite{Logan} Feynman rules for the vertices 
involving $W_L^\pm$,
$W_H^\pm$ and the quarks have been given. We repeat them except that 
we introduce  additional $i$ factors in the rules involving the heavy $T$ 
quark that we will discuss below. We have then
\begin{eqnarray}
W_L^{+\mu}\bar u_id_j&=&i
\frac{g_2}{2\sqrt{2}}\;V_{ij}\;\left[1-a\frac{v^2}{f^2}\right]\gamma_{\mu}\;
(1-\gamma_5) \qquad (u_i=u,c) \label{uvertex}\\
W_L^{+\mu}\bar td_j&=&i
\frac{g_2}{2\sqrt{2}}\;V_{tj}\;
\left[1-\left(\frac{1}{2} x^2_L+a\right)\frac{v^2}{f^2}\right]\gamma_{\mu}\;
(1-\gamma_5)\label{tvertex}\\
W_L^{+\mu}\bar Td_j&=&i
\frac{g_2}{2\sqrt{2}}\;V_{tj}\;x_L\;\frac{v}{f}
\gamma_{\mu}\;
(1-\gamma_5)
\label{Tvertex}
\end{eqnarray}
and
\begin{eqnarray}
W_H^{+\mu}\bar u_id_j&=&-i
\frac{g_2}{2\sqrt{2}}\;V_{ij}\frac{c}{s}
\;\left[1+b\frac{v^2}{f^2}\right]\gamma_{\mu}\;
(1-\gamma_5)\qquad (u_i=u,c) \label{uvertex1}\\
W_H^{+\mu}\bar td_j&=&-i
\frac{g_2}{2\sqrt{2}}\;V_{tj}\frac{c}{s}
\;\left[1-\left(\frac{1}{2} x^2_L-b\right)\frac{v^2}{f^2}\right]\gamma_{\mu}\;
(1-\gamma_5)\label{tvertex1}\\
W_H^{+\mu}\bar Td_j&=&-i
\frac{g_2}{2\sqrt{2}}\;V_{tj}\frac{c}{s}
\;x_L\;\frac{v}{f}
\gamma_{\mu}\;
(1-\gamma_5)
\label{Tvertex1}
\end{eqnarray}
where
\be\label{ab}
a=\frac{1}{2} c^2 (c^2-s^2), \qquad b= \frac{1}{2} s^2 (c^2-s^2).
\ee
The $\ord(v^2/f^2)$ corrections to the $W_H^\pm$ couplings, not contained in 
table VIII of \cite{Logan}, follow from equation (A51) of the latter paper.
The Feynman rules for the leptons are given by (\ref{uvertex}) and 
(\ref{uvertex1}) with $V_{ij}=1$.

Here $V_{ij}$ are the usual CKM parameters, denoted by $V_{ij}^{\rm SM}$ in 
\cite{Logan}. They satisfy the usual unitarity relations. In particular we have
\be\label{unitarity}
\lambda_u+\lambda_c+\lambda_t=0, \qquad \lambda_i=V^*_{ib} V_{id}. 
\ee

As seen in (\ref{tvertex}) and (\ref{tvertex1}) at $\ord(v^2/f^2)$ there
is a disparity between the $W_L^{+\mu}(W_H^{+\mu})$ couplings 
of $t$ and  of the lighter quarks $u$ and $c$ to the down quarks. 
This is related to the fact that
in the LH model the elements $V_{ij}$ are generalized to \cite{Logan,QUAD}
\be\label{CKM0}
\hat{V}_{ij} = V_{ij} \qquad \textrm{for} \qquad i = u,c
\ee
and
\be\label{ckmelements}
\hat{V}_{tj} = V_{tj}\left(1- \frac{x_{L}^2}{2}\frac{v^2}{f^2}\right), 
\qquad \hat{V}_{Tj} = V_{tj} \frac{v}{f} x_{L}
\ee 
and include now also the heavy $T$.

We observe that
the $\mathcal{O}(v^2/f^2)$ corrections to $V_{tj}$ in 
(\ref{ckmelements})  
violate the usual CKM unitarity relations like the one in 
(\ref{unitarity}) but the generalized unitarity relation \cite{QUAD}
\be\label{unitarityLH}
\hat\lambda_u+\hat\lambda_c+\hat\lambda_t+\hat\lambda_T=0, \qquad\qquad
\hat\lambda_i=\hat V^*_{ib}\hat V_{id},
\ee
that includes also the heavy $T$
is clearly satisfied at $\mathcal{O}(v^2/f^2)$.

The Feynman rules in (\ref{uvertex})--(\ref{Tvertex1}) are the same as in 
\cite{Logan}  except for the additional $i$ factors in (\ref{Tvertex}) and 
(\ref{Tvertex1}).
The absence of these factors in 
 table VIII of \cite{Logan} is  related to the fact that with the $i$ factors 
present in the fermion mass terms in equation (A43) of that paper
 the parameter $s_{L}$ in (A44) of \cite{Logan} is an 
imaginary quantity and $s_{L}^{2} + c_{L}^{2} = 1$ is not satisfied. 
Redefining appropriately the quark fields, $s_{L}$ changes to 
$i s_{L} = s_{L}^{new}$, and $\left(s_{L}^{new}\right)^{2} + c_{L}^{2} = 1$. 
The factor $i$ is now present in (\ref{Tvertex}) and (\ref{Tvertex1}) 
as it should be.
In the case of box diagrams with a single $T$ exchange the contribution of 
this heavy quark to $\Delta M_i$ and $\varepsilon_K$ has wrong sign 
if $i$ is not present in (\ref{Tvertex}) and 
(\ref{Tvertex1}) and the divergences in box diagrams 
calculated in the unitary gauge do not cancel. 
We will return to this point below. 

\subsection{Determination of the CKM Parameters}
It is of interest to ask whether the presence of the contributions from 
new particles could 
have an effect on the numerical values of the CKM elements not involving 
$t$ that are usually determined in tree level decays.

In order to address this issue we have to study first the muon decay 
that is usually used to measure the Fermi constant $G_F$. It is sufficient 
to look at the tree level and include only $\ord(v^2/f^2)$ corrections.
In the LH model, in addition to the $W^\pm_L$ exchange also the 
$W^\pm_H$ exchange has to be taken into account. The contribution of 
$\Phi^\pm$ is negligible as it is suppressed both by the $v/f$ factors in
the vertices \cite{Logan} and its large mass.

The Feynman rules for the leptons are identical to the ones for the lighter 
quarks except for the CKM factors. Calculating tree level exchange of 
$W^\pm_L$ with 
$\ord(v^2/f^2)$ corrections taken into account and adding to it the tree 
level exchange of 
$W^\pm_H$ without these corrections gives the standard amplitude for the 
muon decay with $G_F$ replaced by
\be\label{GEFF}
G_F^{\rm eff}=G_F \left(1+ c^2s^2\frac{v^2}{f^2}\right), 
\qquad
\frac{G_F}{\sqrt{2}}=\frac{g^2}{8M_{W_L}^2}.
\ee
To this end we have used the formula for $M_{W^\pm_{H}}$ in (\ref{mass}). It 
is $G_F^{\rm eff}$ that should be identified with the $G_F$ usually 
measured in the muon decay. 

With this information at hand we can now calculate the amplitudes 
for the relevant tree level semileptonic decays in the LH model that 
are used to determine the CKM elements. Proceeding as in the case of the 
muon decay and redefining $G_F$ to $G_F^{\rm eff}$ we find that:
\begin{itemize}
\item
The numerical values of all the CKM elements not involving the top quark 
are not modified at this level.
\item
The numerical values of the CKM elements $V_{tb}$, $V_{ts}$, $V_{td}$ 
determined in tree level decays of the top quark to lighter quarks, 
would also lead to the same results as in the SM but this time for 
$\hat V_{tb}$, $\hat V_{ts}$, $\hat V_{td}$ in (\ref{ckmelements}), 
respectively.
\end{itemize}

This exercise shows immediately how the violation of the three generation 
CKM unitarity in the LH model could be in principle discovered by 
experimentalists in  semi-leptonic decays of $t$ to $b$. Measuring 
$V_{tb},$ but not realizing that what is really measured is $\hat V_{tb}$, 
would give 
the value of $V_{tb}$ that is smaller than the true value. This would 
result in the violation of the unitarity relation
\be\label{utmod}
|V_{ub}|^2+|V_{cb}|^2+|V_{tb}|^2=1
\ee 
with the l.h.s smaller than unity. Realizing that $\hat V_{tb}$ and not
$V_{tb}$ has been measured and using (\ref{ckmelements}) to find the true
 value of $V_{tb}$, would allow to satisfy (\ref{utmod}).
\subsection{Unitarity Triangle in the LH Model}
In view of the $\ord(v^2/f^2)$ corrections to the ordinary CKM elements, 
as given in (\ref{ckmelements}), the unitarity relation for the physical 
CKM elements $\hat V_{ij}$ involving only ordinary quarks is no longer 
satisfied
\be
\hat \lambda_u+\hat \lambda_c+\hat \lambda_t \not=0.
\ee
It would appear then that in the LH model the usual analysis of the unitarity
triangle (UT) should be generalized to a unitarity quadrangle based on 
the relation (\ref{unitarityLH}). 
A discussion in this spirit has been presented in a different context 
in \cite{QUAD}.

Here we would like to emphasize that the usual analysis of the UT remains 
still valid in the LH model, provided we use as basic parameters the 
elements $V_{ij}$ that clearly satisfy the unitarity relation 
(\ref{unitarity}). In this formulation the $v^2/f^2$ corrections to 
the CKM elements in (\ref{ckmelements}) are explicitly seen and 
contribute manifestly to various amplitudes and branching ratios that are 
written in terms of $V_{ij}$ and not $\hat V_{ij}$. The effect of 
the $\ord(v^2/f^2)$ corrections in the CKM elements involving the top quark 
 will be then felt 
together with other corrections in the modification of the numerical values of 
the sides and angles of the UT relative to the ones  obtained in the SM.

Clearly, it is possible to proceed differently and express all amplitudes 
and branching ratios in terms of $\hat V_{ij}$ and not $V_{ij}$. 
In this formulation the $\ord(v^2/f^2)$ corrections to the CKM matrix 
elements will be absorbed into $\hat V_{ij}$ and the explicit 
$\ord(v^2/f^2)$ corrections will differ from the ones in the formulation 
in terms of $V_{ij}$. But as the values of $\hat V_{ij}$ differ from 
$V_{ij}$, as seen explicitly in (\ref{ckmelements}), 
the final result for physical quantities will be the same up to corrections 
of $\ord(v^4/f^4)$.

This discussion is fully analogous to the ones of the definition of 
the QCD coupling constant and the definition of parton distributions 
in deep inelastic scattering. 
We are confident that in the context of the LH model, the variables 
$V_{ij}$ are superior to $\hat V_{ij}$ and we will use them in what follows. 
This allows, in particular, to exhibit the impact of LH effects on 
various processes in the ($\bar\varrho$, $\bar\eta$) plane.

\section{GIM Mechanism and Unitary Gauges}\label{sec:GIM}
\setcounter{equation}{0}
\subsection{Preliminaries}
The amplitudes for FCNC processes in the SM and various extensions 
like supersymmetry and models with extra dimensions, are usually 
calculated in the Feynman gauge or $R_\xi$ gauges for the gauge 
bosons. This requires the inclusion of the corresponding Goldstone bosons 
in order to obtain gauge independent result. In models with larger gauge 
groups, that are spontaneously broken down to the SM group, it is more
convenient to work in the unitary gauge, thus avoiding the calculation 
of many  diagrams with Goldstone bosons. On the other hand, due to different 
high momentum behaviour of gauge boson propagators, even box diagrams 
are divergent in this gauge. These divergences must then cancel each other
 after 
the unitarity of the CKM matrix has been used. To our knowledge no 
explicit demonstration of the cancellation of these divergences 
has been presented in the 
literature. We will first illustrate this within the SM and subsequently 
in the LH model where due to the violation of three generation unitarity 
by $\ord(v^2/f^2)$ corrections, the cancellation in question is more involved.

\subsection{The Standard Model}
In the SM the effective Hamiltonian for $B^0_d-\bar B^0_d$ mixing
neglecting QCD corrections can be written before the use of the CKM
unitarity as follows
\be\label{Heff}
H_{\rm eff} (\Delta B=2)=\frac{G_F^2}{16 \pi^2}M^2_{W^\pm_L}\sum_{i,j=u,c,t}
\lambda_i \lambda_j F(x_i, x_j; W_L) (\bar b d)_{V-A} (\bar b d)_{V-A}
\ee
where
\be\label{prefactor}
\lambda_i=V^*_{ib} V_{id},\qquad  x_i=\frac{m_i^2}{M_{W^\pm_L}^2}.
\ee
The functions $F(x_i, x_j; W_L)$ result up to an overall factor from
box diagram with two $W_L^{\pm}$ and two quarks ($i,j$)
exchanges. 

The unitarity of the CKM matrix implies the relation (\ref{unitarity}).
Inserting $\lambda_u=-\lambda_c-\lambda_t$ into (\ref{Heff}) and
keeping only the term proportional to $\lambda_t^2$ one finds
\be\label{Heff2}
H_{\rm eff} (\Delta B=2)=\frac{G_F^2}{16 \pi^2}\;M^2_{W^\pm_L} \;\lambda_t^2\;
S_0(x_t) (\bar b d)_{V-A} (\bar b d)_{V-A},
\ee
where
\be\label{S0}
S_0(x_t)=F(x_t, x_t; W_L)+F(x_u, x_u; W_L)-2 F(x_u, x_t; W_L).
\ee
Similarly the coefficient of $2 \lambda_c \lambda_t$ is given by
\be\label{S0xcxt}
S_0(x_c, x_t)=F(x_c, x_t; W_L)+F(x_u, x_u; W_L)-F(x_u, x_c;
W_L)-F(x_u, x_t; W_L).
\ee
In any $R_{\xi}$ gauge for the $W^\pm_L$ propagator the functions $F$ are
finite but contain $x_i$-independent terms that, when present, would
be disastrous in particular for the evaluation of the $K_L-K_S$ mass
difference $\Delta M_K$ \cite{LG74}. Such terms evidently 
cancel in (\ref{S0}) and
(\ref{S0xcxt}) and in an analogous expression for $S_0(x_c)$, that to
an excellent approximation, is given then by $x_c$, providing the
necessary suppression of $\Delta M_K$ in accordance with experimental
findings.

In the unitary gauge the functions $F$ are divergent quantities with
the divergence given up to an overall $x_i$-independent factor by
\be\label{Fsing}
F_{\rm div}(x_i, x_j;W_L)\sim \frac{1}{\varepsilon} (x_i + x_j + 
\textrm{const.})
\ee
with $\varepsilon$ defined through $D=4-2 \varepsilon$. It is evident
that these singularities cancel in the expressions (\ref{S0}) and
(\ref{S0xcxt}). We have verified that the remaining terms reproduce
the known expressions for $S_0(x_t)$ and $S_0(x_c, x_t)$ that are given in
Appendix B.

For pedagogical reasons it is instructive to demonstrate how these
divergences disappear when the use of the relations (\ref{S0}) and
(\ref{S0xcxt}) is already done at the level of the integrand so that
the use of the dimensional regularization can be avoided altogether. This
is in fact useful when the calculations are done by hand although
immaterial when computer software for analytical
calculations is used.

In the process of the evaluation of the functions $F(x_i, x_j;W_L)$ in
the unitary gauge, two divergent integrals corresponding respectively
to $g_{\alpha \beta} k_{\mu} k_{\nu}$ and $k_{\alpha} k_{\beta}
k_{\mu} k_{\nu}$ factors appear:
\be\label{int}
I_n (x_i, x_j)=\int_0^{\infty} dr \frac{r^{2+n}}{(r+x_i)(r+x_j) (r+1)^2}
\qquad n=1,2.
\ee
Inserting these integrals into (\ref{S0}) results in finite integrals 
\be\label{int2}
I_n^{\rm GIM}(x_t, x_t)=
x_t^2 \int_0^{\infty} dr \frac{r^{n}}{(r+x_t)^2 (r+1)^2}
\qquad n=1,2
\ee
with an analogous result for $I_n^{\rm GIM}(x_c, x_t)$. It is
remarkable that the GIM mechanism reduces the power in the numerator
by two.

\subsection{Littlest Higgs Model}
As discussed above, at  $\ord(v^2/f^2)$ the CKM matrix involving
only the usual three generations of quarks is no longer unitary.
Consequently when $\ord(v^2/f^2)$ corrections to the $W_L^{\pm} \bar
u_i d_j$ vertices are included and only the exchanges of ordinary
quarks are taken into account the functions $S_0(x_t)$ and $S_0(x_c, x_t)$
are divergent at $\ord(v^2/f^2)$ even if the relation
(\ref{unitarity}), still valid at $\ord(1)$ in the LH model, is
used. This leftover divergence is then cancelled by box diagrams involving a
single $T$ quark in place of an ordinary quark. At $\ord(v^4/f^4)$ the
inclusion of box diagrams with two $T$ quarks becomes necessary.

The basic formula that guarantees the cancellation of the 
quadratic divergences in the LH model is the generalized unitarity 
relation  (\ref{unitarityLH}). 
For this relation to be effective in the evaluation of 
the box diagrams and also penguin diagrams it is essential that the 
Feynman rules 
in (\ref{Tvertex}) and (\ref{Tvertex1}) contain the factor $i$, 
that in fact is not present in 
the corresponding rules in Table VIII of \cite{Logan}. We have discussed this point 
in Section 2.
In the case of box diagrams with a single $T$ exchange, the omission of 
this $i$ factor  would give the wrong sign for the $T$ contribution and 
 the divergences coming from diagrams with ordinary quarks would
  not be cancelled. 
We will return to this point when presenting our results in the subsequent 
section.

\begin{figure}[h]
\vspace{0.10in}
\epsfysize=5in
\centerline{
\epsffile{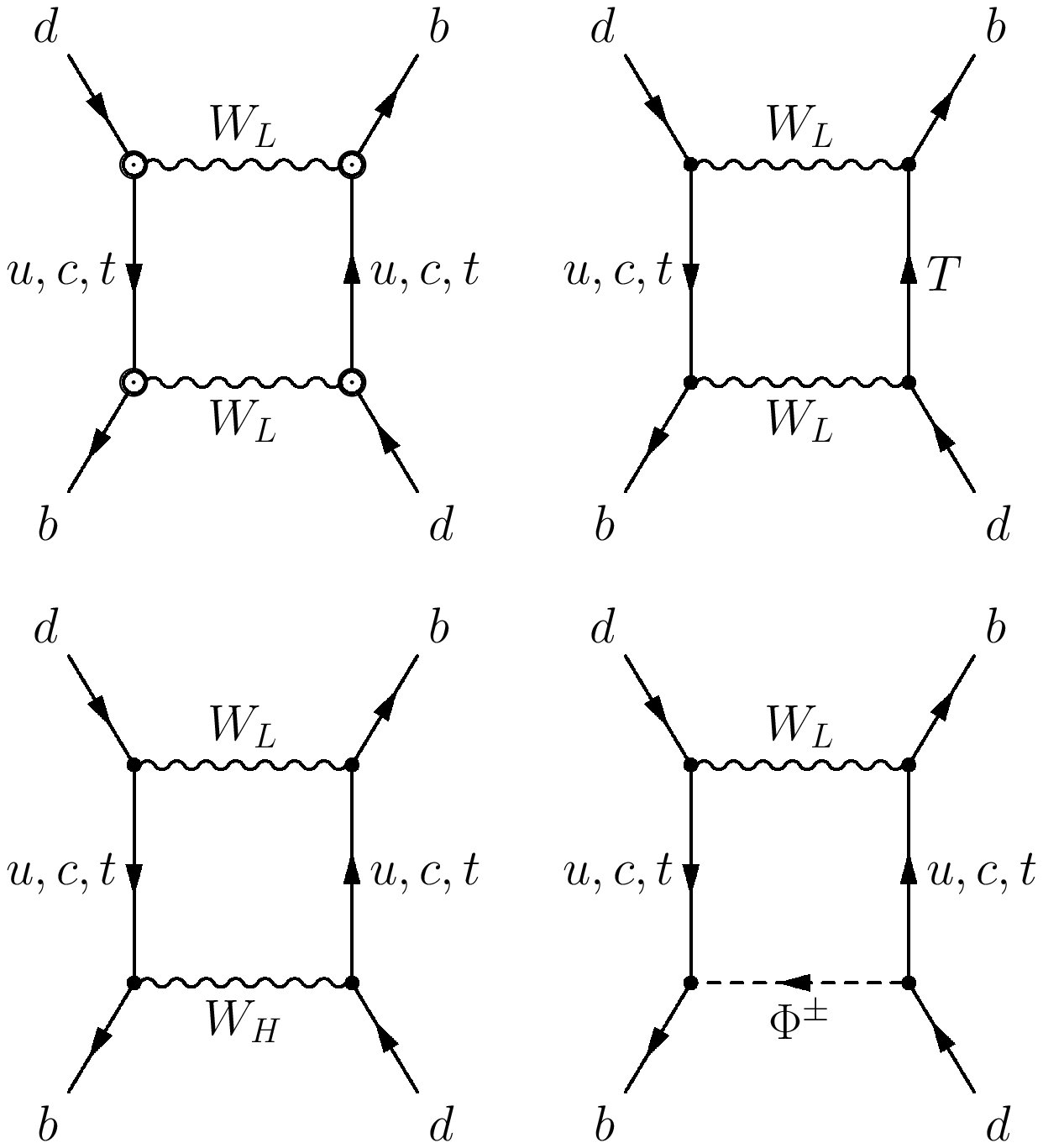}}
\vspace{0.08in}
\caption{Contributing box diagrams at $\ord(v^2/f^2)$}\label{fig:boxes}
\end{figure}

\section{Analytic Results}\label{sec:AR}
\setcounter{equation}{0}
In this section we will present analytic results for the $\ord(v^2/f^2)$ 
corrections to 
$\Delta M_K$, $\Delta M_{d,s}$ and the 
CP-violating parameter $\varepsilon_{K}$ in the LH model.

The effective Hamiltonian for $\Delta S = 2$ transitions 
can be written as follows
\be\label{hamiltonian}
H_{\rm eff} = \frac{G_{F}^{2}}{16 \pi^{2}}M_{W_L^\pm}^{2}
\left[\lambda_{c}^{2} \eta_{1} S_{c} + \lambda_{t}^{2} \eta_{2} S_{t} 
+ 2\lambda_{c}\lambda_{t}\eta_{3}S_{tc}\right] (\bar s d)_{V-A} (\bar s d)_{V-A}, 
\ee
where $\lambda_{i} = V^{*}_{is} V_{id}$. In the case of 
$B_{d}^{0} - \bar{B}_{d}^{0}$ mixing and 
$B_{s}^{0} - \bar{B}_{s}^{0}$ mixing the formula (\ref{Heff2}) applies with 
$S_{0}(x_{t})$ replaced by $\eta_{B} S_{t}$.
The factors $\eta_{i}$ are QCD corrections \cite{HN,BJW90,UKJS} 
to which we will return in 
Section~\ref{SEC6}. 

Using (\ref{hamiltonian}) one obtains the following expressions for 
quantities considered in this paper \cite{Schladming}:
\begin{equation}
\varepsilon_K=C_{\varepsilon} \hat B_K \IM\lambda_t \left\{
\RE\lambda_c \left[ \eta_1 S_c - \eta_3 S_{ct} \right] -
\RE\lambda_t \eta_2 S_t \right\} e^{i \pi/4}\,,
\label{eq:epsformula}
\end{equation}
\begin{equation}
\Delta M_q = \frac{G_ F^2}{6 \pi^2} \eta_B m_{B_q} 
(\hat B_{B_q} F_{B_q}^2 ) M_{W_L}^{2} S_t |\lambda_t|^2, \qquad q=d,s~,
\label{eq:xds}
\end{equation}
where the numerical constant $C_\varepsilon=3.837 \cdot 10^4 $,
$F_{B_q}$ is the $B_q$ meson decay constant, $\hat B_i$ are
non-perturbative parameters 
and $\eta_B$ stands for short distance QCD correction that slightly differs
from $\eta_2$ in (\ref{hamiltonian}) \cite{BJW90,UKJS}.

As discussed in Section 2, it is convenient to work directly 
with $\lambda_{i}$ rather 
than $\hat{\lambda}_{i}$ and include the effects of the corrections to 
the CKM matrix in the functions $S_{i}$ and $S_{ij}$.
We decompose therefore the functions $S_{i}$ and $S_{ij}$ into known SM 
contributions and the corrections coming from new particles in the LH model 
as follows
\begin{eqnarray}
S_{t}\,\, & = & S_{0}(x_{t}) + \Delta S_{t},\\
S_{ct}\, & = & S_{0}(x_{c}, x_{t}) + \Delta S_{ct},\\
S_{c}\,\, & = & S_{0}(x_{c}) + \Delta S_{c}.
\end{eqnarray}
The diagrams contributing to the functions $\Delta S_{i}$ at $\ord(v^2/f^2)$ 
are shown in 
Fig.~\ref{fig:boxes}. The circles around the vertices in the first 
diagram  that involves only SM particles indicate 
$\mathcal{O}(v^2/f^2)$ corrections to the $W_{L}^{\pm}$ vertices. 
Explicit expressions  are given in (\ref{uvertex}) and (\ref{tvertex}). 
No such corrections have to be included in the last two diagrams with 
$W^\pm_{H}$ and $\Phi^{\pm}$, exchanges because of the large masses 
of these particles. The case of the second diagram with $T$ is different 
due to the non-decoupling of $T$. We will return to this point in the next 
section.

This discussion shows that the contribution of the scalars $\Phi^{\pm}$ is 
much smaller than the remaining contributions as the last diagram 
in Fig.~\ref{fig:boxes}
 is suppressed by both $v^2/f^2$ in the vertices involving 
$\Phi^{\pm}$ and the large mass $M_{\Phi^{\pm}}$. 
We have confirmed this expectation through an explicit calculation. 
We therefore omit this contribution in what follows but for completeness 
give the analytic expression for it in  Appendix A.

Similarly the $\ord(v^2/f^2)$ corrections to the third diagram, that 
involve necessarily also a single $T$ exchange like in the second diagram, 
give contributions that can be 
neglected in comparison with the first three diagrams. For 
completeness we give the analytic expression for these corrections in 
 Appendix A.

The expressions for $\Delta S_{i}$ and $\Delta S_{ij}$ that are obtained 
from the first three diagrams of Fig.~\ref{fig:boxes} are then given as 
follows:
\begin{eqnarray}
\Delta S_{t}\, & = & 
-4 \frac{v^2}{f^2} \left[a S_{0}(x_{t}) + 
\frac{1}{2} x_{L}^{2} P_{1}(x_{t}, x_{T})\right] + 
2 \frac{c^2}{s^2} P_{3}(x_{t}, y) \label{tcorrection} \\
\Delta S_{ct} & = & -4 \frac{v^2}{f^2} 
\left[a S_{0}(x_{c}, x_{t}) + 
\frac{1}{4} x_{L}^{2} P_{2}(x_{c}, x_{t}, x_{T})\right] + 
2 \frac{c^2}{s^2} P_{4}(x_{c}, x_{t}, y) \label{ctcorrection}
\end{eqnarray}
with $\Delta S_{c}$ obtained from (\ref{tcorrection}) through the 
substitution $t \to c$ and setting $x_L = 0$. 
The parameters $x_{L}$ and $a$ are defined in (\ref{XL}) and (\ref{ab}), 
respectively.
Moreover
\be
x_{i} = \frac{m_{i}^{2}}{M_{W_L^\pm}^{2}}, 
\qquad y = \frac{M_{W_H^\pm}^{2}}{M_{W_L^\pm}^{2}}.
\ee 

Explicit expressions for the functions $S_0$ and 
$P_{i}$ are given in Appendix B. It turns out that in the range of
parameters considered, the four functions involved can be approximated 
within an excellent accuracy by 
\begin{eqnarray}
P_{1}(x_{t}, x_{T}) \;\;\;\;\;& = & -\frac{x_t}{4} (\log x_T -1.57) \label{AP1} \\
P_{2}(x_{c}, x_{t}, x_{T}) & = & -\frac{x_c}{4} (\log x_T + 0.65) \label{AP2} \\
P_{3}(x_{t}, y)\;\;\;\;\;\;\; & = & \frac{x_t}{y} \label{AP3} \\
P_{4}(x_{c},x_t, y)\;\; & = & \frac{x_c}{y} \log\frac{x_t}{x_c} \label{AP4},
\end{eqnarray}
where the numerical factors correspond to $m_t=168.1~{\rm GeV}$ and
$m_c=1.3~{\rm GeV}$.
However, in our numerical analysis, we will use the exact expressions.

The expressions for the functions $P_{i}$  in terms of 
the functions $F$ resulting from individual diagrams  
are given as follows:
\begin{eqnarray}
\hspace{-1cm}& & P_{1}(x_{t}, x_{T}) = 
F(x_{t}, x_{t}; W_{L}) + 
F(x_{u}, x_{T}; W_{L}) - 
F(x_{t}, x_{T}; W_{L}) - 
F(x_{u}, x_{t}; W_{L}) \label{P1} \\
\hspace{-1cm}& & P_{2}(x_{c}, x_{t}, x_{T})  =
F(x_{c}, x_{t}; W_{L}) + 
F(x_{u}, x_{T}; W_{L}) - 
F(x_{c}, x_{T}; W_{L}) - 
F(x_{u}, x_{t}; W_{L}) \label{P2} \\
\hspace{-1cm}& & P_{3}(x_{t}, y)  = 
F(x_t,x_t, y; W_L,W_H) + 
F(x_u, x_u,y; W_{L},W_H) - 
2 F(x_{t}, x_{u},y ; W_{L},W_H) \label{P3} \qquad \; \\
\hspace{-1cm}& & P_{4}(x_{c},x_t, y)\, =
F(x_c,x_t, y; W_L,W_H) + 
F(x_u, x_u,y; W_{L},W_H) \nonumber \\
\hspace{-1cm}& & \qquad \qquad \qquad \; - F(x_{c}, x_{u},y ; 
W_{L},W_H) - F(x_{t}, x_{u},y ; W_{L},W_H) \label{P4}
\end{eqnarray}

As discussed in Section 3 each contribution in 
(\ref{P1})-- (\ref{P4}) is divergent in the unitary gauge 
but these divergences are absent in $P_{i}$. 
For this to happen the signs in front of the functions having 
the argument $x_{T}$ must be as given above.
This is only achieved with the $i$ factor in the  rules (\ref{Tvertex})
and (\ref{Tvertex1}). 
Removing $i$ from these rules would imply opposite signs in front of 
the functions involving $T$ in (\ref{P1})--(\ref{P2}) and no 
cancellation of 
divergences. This is evident from (\ref{Fsing}). 

The results in (\ref{tcorrection}) and (\ref{ctcorrection})
do not include the correction related to $G_F$ that has been given in
(\ref{GEFF}). Rewriting (\ref{hamiltonian}) in terms of $G_F^{\rm eff}$ 
results in the replacement
\be\label{aeff}
a \to a_{\rm eff}=a+\frac{1}{2}c^2s^2=\frac{1}{2}c^4.
\ee 
This correction slightly suppresses the enhancements of $S_t$ and $S_{ct}$.

The formulae (\ref{tcorrection}) - (\ref{aeff}) 
 and the analytic expressions for the 
functions $P_{i}$ in 
 Appendix B are the main results of this section.

\boldmath
\section{Non-Decoupling Effects of the Heavy $T$}\label{DEC}
\setcounter{equation}{0}
\unboldmath
\subsection{Preliminaries}
In the previous section we have seen that box diagrams including
simultaneously either $W^\pm_H$ or $\Phi^\pm$ and  explicit 
$\ord(v^2/f^2)$ corrections in the vertices could be neglected for all
practical purposes. Effectively they are $\ord(v^4/f^4)$ with the 
additional suppression factor $\ord(v^2/f^2)$ coming from the heavy 
gauge boson or scalar propagator.

This rule does not apply to the box diagram with the single $T$ exchange in
Fig.~\ref{fig:boxes}. Indeed as seen in (\ref{AP1}) the contribution of this diagram 
increases 
logarithmically with $x_T$, rather than being suppressed by a heavy $T$
quark propagator.

In order to understand this particular behaviour of diagrams involving $T$
let us recall the known fact,
that in the SM
the FCNC processes 
are dominated 
by the contributions of top quark exchanges in box and penguin diagrams 
\cite{Schladming,IL,BSS}. 
This dominance originates in the large mass $m_t$ of the top quark and in 
its non-decoupling from low energy observables due to the corresponding Yukawa 
coupling that is proportional to $m_t$. In the evaluation of box and
penguin diagrams in the Feynman-t'Hooft gauge this decoupling is realized
through the diagrams with internal fictitious Goldstone boson and top quark
exchanges. The couplings of Goldstone bosons to the top quark, being
proportional to $m_t$, remove the suppression of the diagrams in question 
due to top quark propagators so that at the end the box and penguin diagrams
increase with increasing $m_t$. In the unitary gauge, in which fictitious
Goldstone bosons are absent, this behaviour originates from the 
longitudinal $(k_\mu k_\mu/M^2_W)$ component of the $W^\pm$--propagators.

In particular, in the case of $B^0_{d,s}-\bar B^0_{d,s}$ mixing and 
$\varepsilon_K$ discussed here, the function  
$S_0(x_t)$ in (\ref{S0}) has the following large $m_t$ behaviour
\be\label{Sasym}
S_0(x_t) \to \frac{x_t}{4}.
\ee 
Yet, with $x_t\approx 4.4$, this asymptotic formula is a  very poor 
approximation 
of the true value $S_0(x_t)=2.42$. 

In the 
case of the Littlest Higgs model, the corresponding variable $x_T$ is 
at least $400$ and 
the asymptotic formula (\ref{AP1}) is an excellent approximation of the 
exact expression for $P_1(x_t,x_T)$. The question then arises, whether 
this formula is an adequate description of the large $m_T$ limit 
that as seen in (\ref{mass}) at fixed $f/v$ corresponds to $x_L\approx 1$. As 
seen in (\ref{lambdarel}) in this limit the Yukawa coupling $\lambda_1$ 
becomes large implying non-decoupling of $T$. Here we want to point out 
that in this limit also the $\ord(v^4/f^4)$ contributions involving
 the $T$ quark represented dominantly by box diagrams with two $T$ 
exchanges must also be taken into account.
In fact for $x_L\ge 0.95$ these $\ord(v^4/f^4)$ corrections turn out to be the
dominant correction in the LH model to the SM result for $S_t$.
A short summary of the results obtained here appeared very recently 
in \cite{DEC05}. Here we present the details of
 these investigations.
\boldmath
\subsection{Box Diagrams with Two $T$ Exchanges}
\unboldmath 
Returning to the results of the previous section, let us 
note that all  contributions calculated there have a characteristic linear
behaviour in $x_t$ that signals the non-decoupling of the ordinary top quark.
However, the corresponding non-decoupling of $T$ is only logarithmic. 
This is related 
to the fact that with the  $W^\pm_L \bar T d_j$ coupling being $\ord(v/f)$ 
only box diagrams with a single $T$ exchange (see Fig.~\ref{fig:boxes}) 
contribute at 
$\ord(v^2/f^2)$. Similarly to the SM box diagrams with a single $t$ 
exchange, that increase as $\log x_t$, the $T$ contribution in the 
LH model increases only as $\log x_T$.

\begin{figure}[h]
\vspace{0.10in}
\epsfysize=2.1in
\centerline{
\epsffile{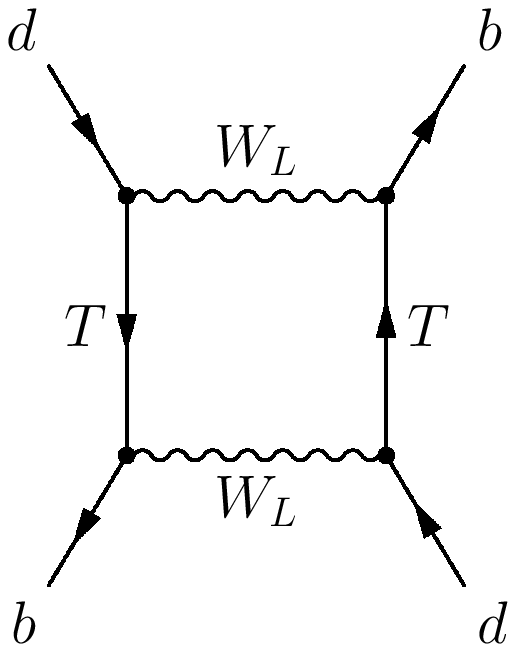}}
\vspace{0.08in}
\caption{The dominant box diagram at $\ord(v^4/f^4)$}\label{fig:TTboxes}
\end{figure}

Yet, as discussed in \cite{DEC05},
 for $x_L\approx 1$ at fixed 
$v/f$, the $\log x_T$ behaviour of the $T$ contribution found in the previous
Section  does not give a proper
description of the non-decoupling of $T$. Indeed in this limit also the
box diagram with two $T$ exchanges given in Fig.~\ref{fig:TTboxes}
 has to be considered. 
Although formally $\ord(v^4/f^4)$, this contribution increases linearly with
$x_T$ and with $x_T=\ord(f^2/v^2)$  constitutes effectively
an $\ord(v^2/f^2)$ contribution.

In order to include the box diagram with two $T$ exchanges in our analysis 
one also has to calculate  
the $\ord(v^4/f^4)$ corrections from
the first two diagrams in Fig.~\ref{fig:boxes},
that have to be taken into account in
order to remove the divergences characteristic for a unitary gauge calculation
and for the GIM mechanism \cite{GIM} to become effective.

To this end the rules for the vertices in (\ref{tvertex}) and 
(\ref{Tvertex}) have to be extended 
to the next order in $v/f$. Keeping only terms involving $x_L$, 
that are relevant 
for this discussion, let us write then 
\begin{eqnarray}
W_L^{+\mu}\bar td_j&=&i
\frac{g_2}{2\sqrt{2}}\;V_{tj}\;
\left[1-\frac{1}{2} x^2_L\frac{v^2}{f^2}   
+\left(d_1+\frac{1}{2} a x^2_L\right)\frac{v^4}{f^4} \right]\gamma_{\mu}\;
(1-\gamma_5)\label{tvertnew}\\
W_L^{+\mu}\bar Td_j&=&i
\frac{g_2}{2\sqrt{2}}\;V_{tj}\;x_L\;\frac{v}{f}
\left[1+\left(d_2 - a\right)\frac{v^2}{f^2}\right]
\gamma_{\mu}\; (1-\gamma_5),
\label{Tvertnew}
\end{eqnarray}
where the coefficients $d_1$ and $d_2$ could in principle be found by 
extending the analysis in \cite{Logan} to include $\ord(v^4/f^4)$
corrections. 
Fortunately,
in order to find the dominant $\ord(v^4/f^4)$ correction coming from 
the diagram in Fig.~\ref{fig:TTboxes}, 
the detailed knowledge of $d_1$ and $d_2$ turns out
to be unnecessary. The reason is that in order to cancel all 
divergences or equivalently to satisfy the generalized unitarity 
relation in (\ref{unitarityLH}) at $\ord(v^4/f^4)$, these two coefficients 
have to be related to each other as follows
\be\label{reld1d2}
d_1=-d_2 x_L^2-\frac{x_L^4}{8}~.
\ee 

Indeed with (\ref{ckmelements}) generalized to
\be\label{ckmnew}
\hat{V}_{tj} = 
V_{tj}\left(1- \frac{x_{L}^2}{2}\frac{v^2}{f^2} +d_1\frac{v^4}{f^4}\right), 
\qquad \hat{V}_{Tj} = 
V_{tj} \frac{v}{f} x_{L}\left(1+d_2 \frac{v^2}{f^2}\right)
\ee
the relation (\ref{unitarityLH}) is only satisfied at $\ord(v^4/f^4)$,
provided $d_1$ is related to $d_2$ as in (\ref{reld1d2}).

Using this relation, the result for the sum of the diagram in 
Fig.~\ref{fig:TTboxes} 
and the $\ord(v^4/f^4)$ corrections from
the first two diagrams in Fig.~\ref{fig:boxes}   
can be written as
\be\label{DTTS}
(\Delta S_t)_{TT}=\frac{v^4}{f^4}
\left[x_L^4 P_{TT}(x_t,x_T)-4 (d_2 -2a) x_L^2 P_1(x_t,x_T)\right]
\ee
with $P_1(x_t,x_T)$ given already in (\ref{AP1}) and $P_{TT}(x_t,x_T)$ given 
simply as follows
\be\label{PTTm}
P_{TT}(x_t,x_T)=
F(x_T, x_T; W_L)+F(x_t, x_t; W_L)
-{2} F(x_t, x_T; W_L)~.
\ee
The meaning of the functions $F(x_i, x_j; W_L)$ is as in the previous 
section. Exact formula for $P_{TT}(x_t,x_T)$ is given in the Appendix B. 

As $P_1(x_t,x_T)$ increases only logarithmically with $x_T$, the second term
in (\ref{DTTS}) is a genuine $\ord(v^4/f^4)$ correction and can be safely 
neglected.
On the other hand, the first term, that is independent of $d_i$,
 gives for $x_L \approx 1$
\be\label{DTTSa}
(\Delta S)_{TT}\approx\frac{v^4}{f^4} x_L^4 \frac{x_T}{4}
=
\frac{v^2}{f^2} \frac{x_L^3}{1-x_L} \frac{x_t}{4}~.
\ee
Formula (\ref{DTTSa})  represents for $x_L>0.85$ and $f/v\ge 5$
 the exact expression given in Appendix B to within $3\%$ and becomes
rather accurate for $x_L>0.90$ and $f/v\ge 10$.

In fact the result in (\ref{DTTSa}) can easily be understood.
In the limit of a very large $x_T$ 
it turns out to be a good approximation to 
evaluate $P_{TT}(x_t,x_T)$ with
$x_t=0$. In this case (\ref{PTTm})  reduces to
$S_0(x_t)$ in (\ref{S0}) with 
$x_t$ replaced by $x_T$ and $x_u$ by $x_t$. 
The factor ${x_t}/{4}$ in (\ref{Sasym}) 
is then replaced by $x_T/4$ as seen in (\ref{DTTSa}).

The formula (\ref{DTTSa}) 
 and the exact expression for $P_{TT}(x_t,x_T)$  
in   Appendix B is the main result of this section.

\section{Numerical Analysis}\label{sec:numerics}
\setcounter{equation}{0}
\subsection{Input Parameters}
We will now evaluate the size of the contributions 
$\Delta S_{c}$, $\Delta S_{t}$ and $\Delta S_{ct}$ as given in 
Section \ref{sec:AR}. 
 To this end we 
use the values of $m_t$ and $M_{W_L^\pm}$ in (\ref{ninput})
 and the following ranges 
for the three new parameters
\be
5 \leq {f}/{v} \leq 20, \quad 0 < x_{L} \leq 0.95, \quad 
0.2 \leq s \leq 0.8.
\ee
This parameter space is larger than the one allowed by other processes 
\cite{Logan}-\cite{PHEN6} which typically imply $f/v\ge 10$ or even higher. 
But we want to demonstrate that even for $f/v$ as low as $5$, the corrections
from LH contributions in this range of parameters, except for $x_L > 0.80$, 
are at most $20\%$.

\boldmath
\subsection{The Size of the Corrections ($x_L\le 0.8$)}
\unboldmath
Let us first analyze the size and the relative importance of the 
explicit $\mathcal{O}(v^2/f^2)$ corrections 
in (\ref{tcorrection}) and of the $W_{H}^{\pm}$ contribution represented by the 
last term in (\ref{tcorrection}). We denote them by $\Delta S_1$ and 
$\Delta S_2$, respectively. Using the formulae (\ref{AP1}) and (\ref{AP3}) 
we have
\be\label{DS1} 
\Delta S_1=\frac{v^2}{f^2} \left[ \frac{x_L^2}{2} x_t (\log x_T -1.57) -
4 a S_0(x_t)\right]
\ee
\be\label{DS2}
\Delta S_2= 2\frac{v^2}{f^2} x_t (1-s^2)^2,
\ee
where we have used (\ref{mass}).

\begin{figure}[ht]
\vspace{0.10in}
\epsfysize=4in
\centerline{
\epsffile{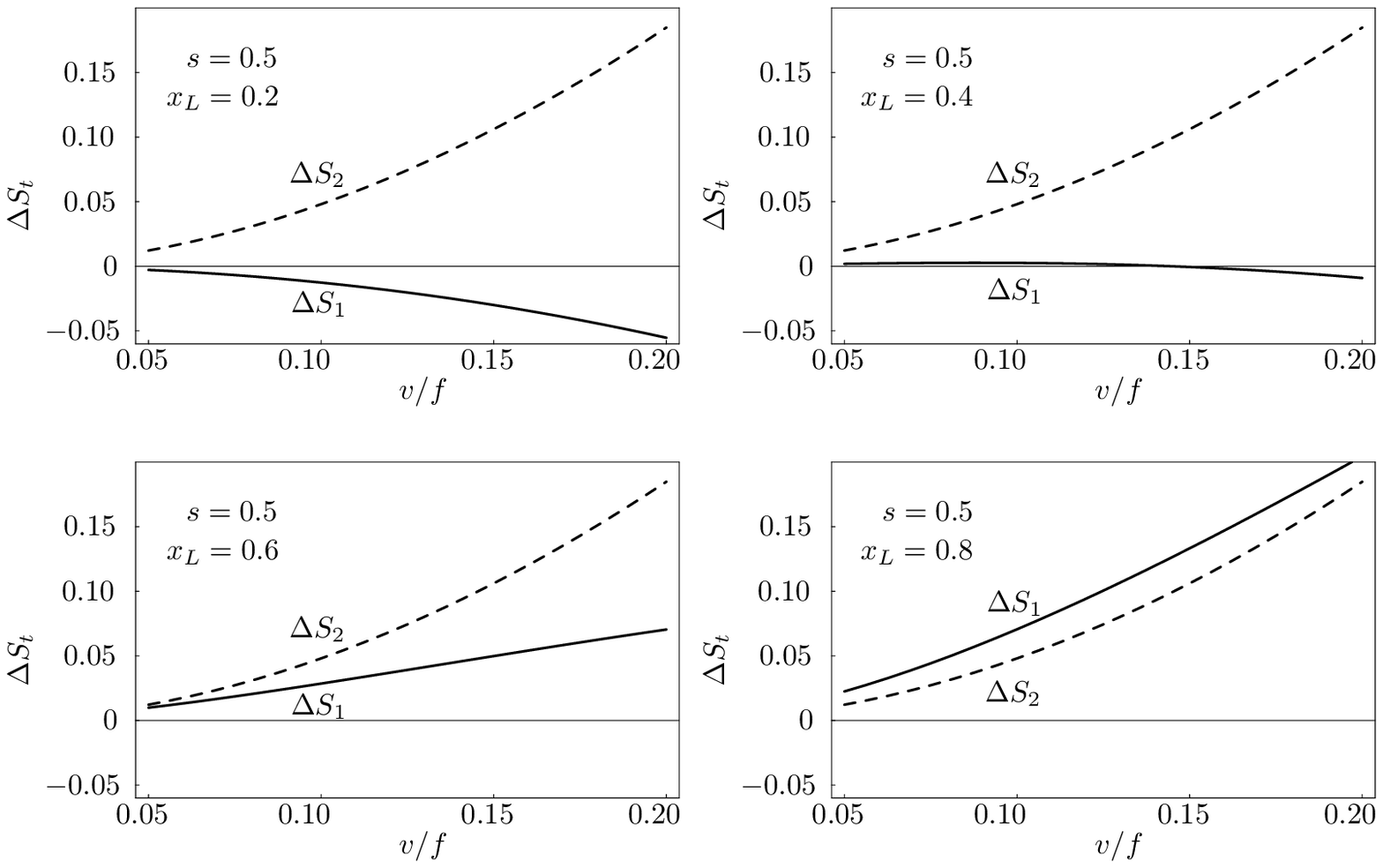}}
\vspace{0.08in}
\caption{The Anatomy of Leading Contributions for $x_L\leq 0.8$.}\label{fig:deltas}
\end{figure}

In Fig.~\ref{fig:deltas} we show $\Delta S_1$ and $\Delta S_2$ 
as functions of ${v}/{f}$ for different values of $x_{L}$ and $s = 0.5$.
For this value of $s$, $\Delta S_2$ is significantly more important 
than $\Delta S_1$ except for the largest $x_L$, where they are comparable 
and have the same sign. The inspection of the formulae (\ref{DS1}) and 
(\ref{DS2}) shows that for larger $s$  and largest $x_L$, 
$\Delta S_1$ can be more important than $\Delta S_2$, while for smaller $s$ 
the dominance of $\Delta S_2$ increases.

\begin{figure}[ht]
\vspace{0.10in}
\epsfysize=4in
\centerline{
\epsffile{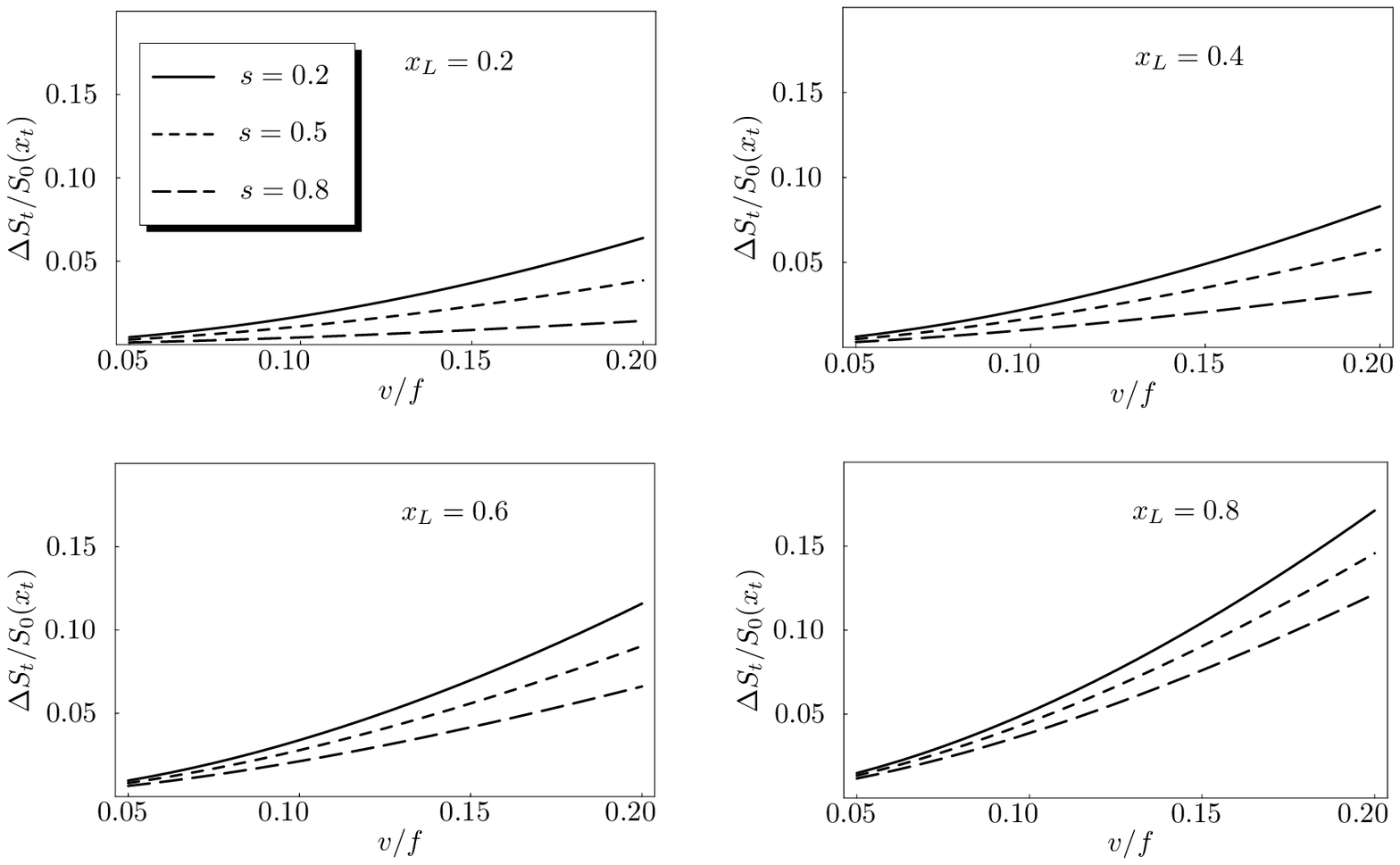}}
\vspace{0.08in}
\caption{$\Delta S_t/S_0(x_t)$ versus $v/f$ for different $s$ and $x_L$}
\label{fig:sinus}
\end{figure}

In Fig.~\ref{fig:sinus} we show the ratio 
$\Delta S_{t}/ S_{0}(x_{t})$ as a function of ${v}/{f}$ 
for different values of $x_{L}$ and $s$. 
In this plot we have taken into account the correction in (\ref{aeff}).
This figure can be compared with the Fig. 2 of \cite{IND1} demonstrating that 
the corrections to the SM expectations in the LH model found by us differ 
both in magnitude and sign from those found in \cite{IND1}. 
We have also calculated the ratios 
$\Delta S_{ct}/ S_{0}(x_{c}, x_{t})$ and 
$\Delta S_{c}/ S_{0}(x_{c})$ to find that, in the whole range of 
parameters considered, they are below $0.03$ and $0.02$, respectively.
In view of hadronic uncertainties in the evaluation of
 $\Delta M_{K}$ and $\varepsilon_{K}$ that amount to at least 
$10 \%$, the corrections $\Delta S_{ct}$ and $\Delta S_{c}$ can be neglected 
for all practical purposes.

We conclude therefore that
\begin{itemize}
\item The corrections from new contributions to $\Delta M_{K}$, 
that is governed by $S_{0}\left(x_{c}\right)$, can be safely neglected.
\item In the case of $\varepsilon_{K}$, $\Delta M_{d}$ and $\Delta M_{s}$ 
the new physics contributions enter to an excellent approximation universally
only the function $S_{t}$ but the observed enhancement in the range of 
parameters considered is by at most $20\%$.
\end{itemize}

\boldmath
\subsection{The Region $x_L\approx 1$}
\begin{figure}[ht]
\vspace{0.10in}
\epsfysize=2in
\centerline{
\epsffile{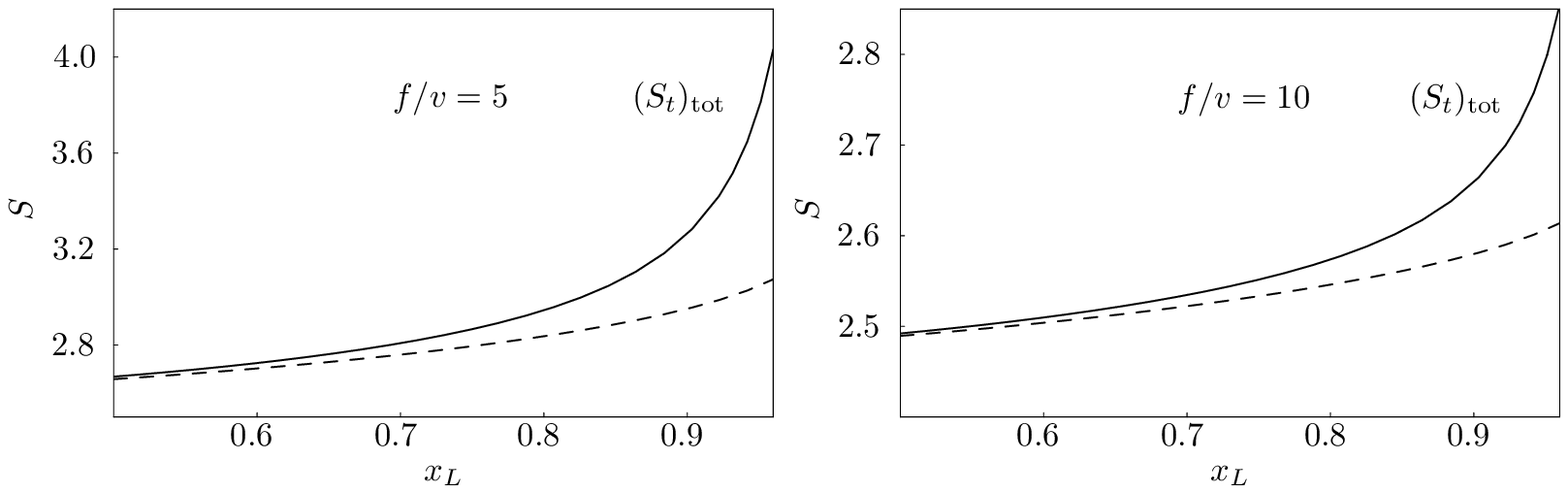}}
\vspace{0.08in}
\caption{$(S_t)_{\rm tot}$ (solid line) versus $x_L$ 
  for $s=0.2$ and $f/v=5,~10$. The dashed line represents the result 
without $TT$ contribution. In the SM $S=S_0(x_t)=2.42$.}
\label{fig:TTResult}
\end{figure}

\unboldmath
Let us next investigate the size of corrections for $x_L\ge 0.8$ where the box
diagrams with two $T$ exchanges become important. 
In Fig.~\ref{fig:TTResult} we show (solid line)
\be\label{Stot}
(S_t)_{\rm tot}=S_0(x_t)+\Delta S_t+(\Delta S)_{\rm TT}
\ee 
as a function of $x_L$ for $f/v=5$ and $f/v=10$ and $s=0.2$.
The comparison with the results for $\Delta S_t$ obtained by 
means of the formulae of Section~\ref{sec:AR} (dashed lines) shows that 
for $x_L\ge 0.8$ 
the box diagrams with two $T$ exchanges cannot be neglected and in fact for 
$x_L\ge 0.95$ they constitute the dominant correction. The total 
correction to the SM result amounts for $x_L=0.95$ and $f/v=5,10$ to 
$56\%$ and $15\%$, respectively. 
Additional numerical results can be found in \cite{DEC05}.

\subsection{Implications}
The enhancement of the function $S_t$ relative to $S_{0}(x_{t})$, 
without the introduction of new operators and new complex phases beyond 
the KM phase is characteristic for all known models with minimal 
flavour violation (MFV) \cite{MFV} like the MSSM at low $\tan \beta$ 
\cite{EPE00}, 
and models with a single universal extra dimension \cite{BSW}. 
Consequently, with the size of corrections found here for $x_L\le 0.80$,
 it will be difficult to distinguish the Littlest Higgs model 
from other MFV models on the basis of particle-antiparticle mixing and 
$\varepsilon_{K}$ alone.
On the other hand for $x_L\ge 0.90$ a distinction could in principle be possible.

The size of the enhancement of $S_{t}$ found here is for $x_L\le 0.80$ 
comparable to the one 
present in models with a single universal extra dimension \cite{BSW} but 
for $x_L>0.90$ it is significantly larger and comparable
with
the maximal enhancements still allowed in the MSSM at low 
$\tan \beta$ \cite{EPE00}. 
Taking into account that in the LH model there are no new 
complex phases and the asymmetry $a_{\Psi K_{\textrm{s}}}$ measures the 
true angle $\beta$ in the UT, the enhancement of $S_t$ with respect to 
$S_{0}(x_{t})$ in the SM implies through the  formulae 
for $\varepsilon_{K}$ and $\Delta M_{d,s}$ in (\ref{eq:epsformula}) and 
(\ref{eq:xds}) 
\be
\left(|V_{td}|\right)_{LH} < \left(|V_{td}|\right)_{SM}, 
\quad \left(R_{t}\right)_{LH} < \left(R_{t}\right)_{SM}, 
\quad \gamma_{LH} < \gamma_{SM}, 
\quad \left(\Delta M_{s}\right)_{LH} > \left(\Delta M_{s}\right)_{SM}
\ee
with $R_t$ being the length of one of the sides of the UT.
However, the suppressions and enhancements of these four quantities are 
at most by $15\%$ for $v/f\le 0.1$ as required by other processes. Such
effects will be very difficult to detect unless 
the theoretical uncertainties in the relevant hadronic uncertainties 
will be decreased well below $5 \%$. We have for instance
\be
\frac{\left(R_{t}\right)_{LH}}{ \left(R_{t}\right)_{SM}}=
\sqrt{\frac{S_t}{S_0(x_t)}}, \qquad 
\frac{\left(\Delta M_{s}\right)_{LH}}{\left(\Delta M_{s}\right)_{SM}}=
\frac{S_t}{S_0(x_t)},
\ee
where we have set the QCD corrections in the LH model and the SM to be equal 
to each other. We will discuss this issue in the next section.

In view of these findings, there is really no useful bound on $f$ coming 
from the processes considered here when $x_L\le 0.80$. 
A rough bound on $f$ in this case turns out to be
\be\label{fbound}
f \ge 1~{\rm TeV}
\ee
that is  weaker than the bound on $f$ of $2-4$ TeV found in 
 in analyses of electroweak precision observables 
\cite{Logan,PHEN1,PHEN2,PHEN3}.
Only in the case of $x_L\ge 0.90$, some significant restrictions on 
the parameter 
space $(x_L,f/v)$ can in principle be obtained, 
provided the hadronic uncertainties 
present in $\Delta M_{s,d}$ will be considerably decreased. We refer to 
\cite{DEC05} for more details. 

\section{Comments on QCD Corrections}\label{SEC6}
\setcounter{equation}{0}
Until now our discussion assumed that the QCD factors $\eta_{i}$ in 
(\ref{hamiltonian}) and (\ref{eq:xds}) were the same for the SM and the 
LH model. 
In fact in the leading logarithmic approximation (LO) this would be true 
if all new heavy particles where $\mathcal{O}(m_{t})$. 
Indeed at the LO what matters is only the renormalization group evolution 
of the $\Delta F = 2 \quad \left(V - A\right) \otimes \left(V - A\right)$ 
operator that for scales below $\mu_{t} = \mathcal{O}(m_{t})$ is 
the same in the SM and the LH model. At the next-to-leading level explicit 
$\mathcal{O}(\alpha_{s})$ corrections to the diagrams of 
Fig.~\ref{fig:boxes} enter, making the $\eta_{i}$ factors in the SM and 
in the LH model  differ by a small amount \cite{HN,BJW90,BBL}. 

However integrating out simultaneously the heavy $T$, $W^\pm_{H}$ and 
$\Phi^{\pm}$ and the significantly lighter $t$ and $W^\pm_{L}$, as we have 
done by calculating the diagrams of Fig.~\ref{fig:boxes}, is certainly a 
rough approximation. Indeed assuming that $T$, $W^\pm_{H}$ and $\Phi^{\pm}$ 
have masses of $\mathcal{O}(f)$, the correct inclusion of 
QCD corrections and summation of large logarithms would require the 
removal of  $T$, $W^\pm_{H}$ and $\Phi^{\pm}$ as explicit degrees of freedom 
at 
$\mu_{f} = \mathcal{O}(f)$ and of ${t}$ and $W_{L}^\pm$ 
at $\mu_{t} = \mathcal{O}(m_{t})$. 
Our experience with the calculations of $\eta_{1}$ and $\eta_{3}$ 
\cite{HN}
tells us 
that in the range $\mu_{t} < \mu < \mu_{f}$ new operators would enter 
the renormalization group analysis even in the case of the term 
$\lambda_{t}^{2}$. Only after $W^\pm_{L}$ and $t$ have been integrated out 
at $\mu_{t}$, would the only operator left in the effective theory 
 be the one in (\ref{hamiltonian}).

A renormalization group analysis for scales $\mu_{t} < \mu < \mu_{f}$ is 
clearly  involved and certainly far beyond the scope of our paper. 
Moreover, in view of the smallness of the corrections found by us, it is 
difficult to justify such an involved analysis. On the other hand the 
experience with the calculations of QCD corrections to the quantities 
considered within the SM \cite{BJW90,HN,BBL} 
indicates that the inclusion of renormalization group effects in the 
range $\mu_{t} < \mu < \mu_{f}$ would likely suppress the LH corrections 
further. However, without a detailed analysis we cannot prove it at present.
As for $\mu \ge m_t$, $\alpha_s$ runs very slowly, the renormalization group
effects in the range $\mu_t\le \mu\le \mu_f$ with $\mu_f=\ord(f)$ are not 
expected to change our main conclusions.

\section{Conclusions}\label{SEC7}
\setcounter{equation}{0}
\vspace*{0.5truecm}

In this paper we have calculated first $\ord(v^2/f^2)$ corrections to the SM
expectations for $\Delta M_K$, $\Delta M_{d,s}$ and $\varepsilon_K$ in 
the Littlest Higgs
model in the case $x_L\le 0.8$. 
The analytic expressions for these corrections are
given in (\ref{tcorrection}) and (\ref{ctcorrection}) and the numerical 
results in Figs.~\ref{fig:deltas} and \ref{fig:sinus}.  Our
main findings for this range of $x_L$ are as follows:

\begin{itemize}
\item
  The dominant new contributions come from box diagrams
       with $(W_L^\pm,W_H^\pm)$ and ordinary quark exchanges 
       that are strictly positive.
\item
The $\ord(v^2/f^2)$ corrections to the usual box diagrams with
two $W_L^\pm$ and ordinary quark exchanges have to be combined with box
diagrams with a single heavy $T$ exchange for GIM mechanism to
work and to cancel the divergences that appear when the
calculation is done in the unitary gauge. These
corrections turn out to be both negative and positive,
dependently on the values of parameters involved, and are smaller than 
those coming
from box diagrams with $(W_L^\pm,W_H^\pm)$  exchanges 
except for $x_L$ approaching $0.8$ when they start to be important.
\item
 The contributions of the heavy scalars $\Phi^\pm$ are
negligible.
\item
The corrections to $\Delta M_K$ are negligible.
\item
   The corrections to $\Delta M_{d,s}$ and $\varepsilon_K$ are positive in 
the full
range of parameters considered. This implies the suppression
of $|V_{td}|$ and of the angle $\gamma$ in the unitarity triangle and an
enhancement of $\Delta M_s$ relative to the SM expectations.
\item
  However even for $f$ as small as 1 TeV, these effects
amount to at most $(15-20)\%$ corrections and decrease below $5\%$
for $f>3-4$ TeV as required by other processes \cite{Logan}--\cite{PHEN6}.
In view of non-perturbative uncertainties in the
quantities considered it will be very difficult to distinguish
LH model from the SM on the basis of particle-antiparticle
mixing and $\varepsilon_K$ alone if $x_L\le 0.8$.
\end{itemize}

Interestingly,
\begin{itemize}
\item
the size of corrections increases for $x_L\approx 1$, where the diagrams 
with two $T$ exchanges become dominant. The relevant expression is given 
in (\ref{DTTSa}) and the numerical results in Fig.~\ref{fig:TTResult}.
 Now the corrections are 
sufficiently large that the distinction from SM expectations for $\vtd$, 
$\gamma$ and $\Delta M_s$ could in principle be possible. 
The corresponding numerical analysis can be found in \cite{DEC05}.
\end{itemize}

Finally,
\begin{itemize}
\item
we have emphasized, that the concept of the unitarity triangle is still useful
in the LH model, in spite of the $\ord(v^2/f^2)$ corrections
to the CKM unitarity involving only ordinary quarks. To this end the 
basic CKM parameters to be used are the uncorrected ones. This should
be useful for future studies of rare decays.
\item
One message is, however, clear: if $\Delta M_s$ will be found
convincingly below the SM expectations, the LH model will be
ruled out independently of the value of $f$.
\end{itemize}

Our results differ significantly from the ones obtained in the published 
version of \cite{IND1},
where a significant suppression of $\Delta M_d$ has been found. 
Meanwhile, the authors identified errors in their calculation and confirmed 
our results of Section \ref{sec:AR}.

It will be interesting to see whether the LH contributions
to theoretically clean rare decays like $K\to\pi\nu\bar\nu$ 
\cite{BSU} will be easier 
to
detect than in quantities considered here. 
This issue is briefly discussed in \cite{DEC05}.
We will present a detailed analysis of rare
decays, that is much more involved, in \cite{BPU05}, where also the 
comparison with the analysis of \cite{IND2} will be given.

\noindent
{\bf Acknowledgements}\\
\noindent
We would like to thank Thorsten Ewerth, Wolfgang Hollik, Heather 
Logan and Felix Schwab for useful discussions. 
Special thanks go to an unknown referee of the first version of our paper 
 who asked us to explain the non-decoupling of $T$ at $\ord(v^2/f^2)$. 
This led us to reconsider our calculation of the function $S$ and to 
include the box diagrams with two $T$ exchanges. In this 
context we thank also Andreas Weiler for useful discussions. 
The work presented here was supported in part by the German 
Bundesministerium f\"ur
Bildung und Forschung under the contract 05HT4WOA/3 and by the German-Israeli
Foundation under the contract G-698-22.7/2002.




\newpage
\begin{appendix}
\section{The Non-Leading Contributions}
\setcounter{equation}{0}
For completeness we give here non-leading contributions to $\Delta S_t$, 
and $\Delta S_{ct}$.

The explicit $\ord(v^2/f^2)$ corrections in the vertices of the third diagram
in Fig.~\ref{fig:boxes} result in
\be\label{CORRt}
(\Delta S_{t})_{W_LW_H} = 
-8 \frac{v^2}{f^2} \frac{c^2}{s^2}
 \left[\tilde a P_3(x_t,y) + 
\frac{1}{2} x_{L}^{2} P_{5}(x_{t}, x_{T},y)\right] 
\ee

\be\label{CORRct}
(\Delta S_{ct})_{W_LW_H} = 
-8 \frac{v^2}{f^2} \frac{c^2}{s^2}
 \left[\tilde a P_4(x_c,x_t,y) + 
\frac{1}{4} x_{L}^{2} P_{6}(x_c,x_{t}, x_{T},y)\right] 
\ee
with $(\Delta S_{c})_{W_LW_H}$ obtained from (\ref{CORRt}) through the 
substitution $t \to c$. 
Here
\be
\tilde a=\frac{a-b}{2}=\frac{(c^2-s^2)^2}{4}
\ee
with $a$ and $b$ defined in (\ref{ab}). 
The correction (\ref{aeff}) contributes here at $\ord(v^4/f^4)$.
The functions $P_{3,4}$ are defined in Section 4 and 
$P_{5,6}$ are defined as follows
\begin{eqnarray}
P_{5}(x_{t},x_T, y) & = &
F(x_t,x_t, y; W_L,W_H) + 
F(x_u, x_T,y; W_{L},W_H) \nonumber \\
& - & F(x_{t}, x_{T},y ; W_{L},W_H)  
- F(x_{t}, x_{u},y ; W_{L},W_H)  \label{P5} \\
P_{6}(x_{c},x_t,x_T,y) & = &
F(x_c,x_t, y; W_L,W_H) - 
F(x_c, x_T,y; W_{L},W_H) \nonumber \\
& - & F(x_{t}, x_{u},y ; W_{L},W_H)
+ F(x_{T}, x_{u},y ; W_{L},W_H)  \label{P6}
\end{eqnarray}
Explicit expressions for $P_{i}$ are given in  appendix B.
We emphasize that these results include $\ord(v^2/f^2)$ corrections 
to both $W^\pm_L$ and $W^\pm_H$ vertices, whereas in \cite{IND1}
 only corrections 
to $W^\pm_L$ vertices have been included. 
In that case $\tilde a=a/2$.

The contribution of the fourth diagram in Fig.~\ref{fig:boxes}
 to $\Delta S_t$ reads

\be\label{Scalar}
(\Delta S_{t})_{W_L\Phi} = 
\frac{1}{2} \frac{v^2}{f^2} 
  P_7(x_t,z)  
\ee
with $P_7$ given by
\be\label{P7}
P_{7}(x_{t}, z)  =  
F(x_t,x_t, z; W_L,\Phi) + 
F(x_u, x_u,z; W_{L},\Phi) - 
2 F(x_{t}, x_{u}, z; W_{L},\Phi)  
\ee
and $z$ defined in (\ref{defs}). In obtaining (\ref{Scalar}) we have 
set the vacuum expectation value $v^\prime$ of the scalar triplet to zero.

We find that in the full range of parameters  given in (\ref{mass})
one has
\be
\frac{(\Delta S_{t})_{W_LW_H}}{S_0(x_t)}\le 3 \cdot 10^{-3}, \quad
\frac{(\Delta S_{ct})_{W_LW_H}}{S_0(x_c,x_t)}\le 3\cdot 10^{-3}, \quad
\frac{(\Delta S_{t})_{W_L\Phi}}{S_0(x_t)}\le 1\cdot 10^{-3}.
\ee
Consequently, all these contributions can be neglected.

\boldmath
\section{The Functions $S_0$ and $P_i$}
\unboldmath
\setcounter{equation}{0}
In the following we list the functions $S_0$ and $P_i$ that 
entered various formulae of our paper. We use
\be\label{defs}
x_i=\frac{m_i^2}{M_{W^\pm_L}^2},\qquad y
=\frac{M_{W^\pm_H}^2}{M_{W^\pm_L}^2}, \qquad
z=\frac{M_{\Phi}^2}{M_{W_L^\pm}^2}.
\ee

\begin{eqnarray}
S_0(x_t)\;\;\;\;\;\;\;\;\;\;\,\,\,\;\;\;&=&\frac{x_t\,( 4 - 11\,x_t + x_t^2) }
   {4\,{( -1 + x_t ) }^2} + 
  \frac{3\,x_t^3\,\log x_t}{2\,{( -1 + x_t ) }^3}\\
\nonumber\\
S_0(x_c, x_t)\;\;\;\;\;\,\;\;\;\;&=&\frac{-3 x_t x_c}
   {4 ( -1 + x_t)( -1 + x_c ) } - 
  \frac{x_t( 4 - 8 x_t + x_t^2 ) x_c \log x_t}
   {4 {( -1 + x_t) }^2
     ( -x_t + x_c ) }\nonumber\\
& + &
  \frac{x_t x_c ( 4-8 x_c +x_c ^2 ) \log x_c}
   {4 {( -1 + x_c ) }^2
     ( -x_t + x_c ) }\\
\nonumber\\
P_{1}(x_{t}, x_{T}) \;\;\;\,\;\;\;\;\,& = & \frac{x_{t}(-4 + 11 x_{t} - x_{t}^{2} + x_{T} - 8 x_{t} x_{T} + x_{t}^{2} x_{T})}{4(-1 + x_{t})^{2}(-1 + x_{T})} + \frac{x_{t} x_{T}(4 - 8 x_{T} + x_{T}^{2}) \log{x_{T}}}{4(x_{t} - x_{T}) (-1 + x_{T})^{2}}
\nonumber \\
& - & \frac{x_{t}(-6 x_{t}^{3} - 4 x_{T} + 12 x_{t} x_{T} - 3
  x_{t}^{2} x_{T} + x_{t}^{3} x_{T}) \log{x_{t}}}{4(-1 +
  x_{t})^{3}(x_{t} - x_{T})}\\
\nonumber\\
P_{2}(x_{c}, x_{t}, x_{T})\;\;\;\; & = & \frac{3 (x_{t} x_{c} - x_{T}
  x_{c})}{4(-1 + x_{t})(-1 + x_{T}) (-1 + x_{c})} + \frac{(4 x_{t}
  x_{c} - 8 x_{t}^{2} x_{c} + x_{t}^{3} x_{c}) \log{x_{t}}}{4(-1 +
  x_{t})^{2}(x_{t} - x_{c})} \nonumber \\
& + & \frac{(4 x_{t} x_{c}^{2} - 4 x_{T} x_{c}^{2} - 8  x_{t} x_{c}^{3} + 8 x_{T} x_{c}^{3} + x_{t} x_{c}^{4} - x_{T} x_{c}^{4}) \log{x_{c}}}{4(x_{t} - x_{c})(x_{T} - x_{c}) (-1 + x_{c})^{2}}\\
& - & \frac{(4 x_{T} x_{c} - 8 x_{T}^{2} x_{c} + x_{T}^{3} x_{c})
  \log{x_{T}}}{4(-1 + x_{T})^{2}(x_{T} - x_{c})} \nonumber \\
\nonumber\\
P_{3}(x_{t}, y)\;\;\;\;\;\;\,\;\;\;\; & = & \frac{x_{t}(-4 x_{t} + x_{t}^{2} + 4 y - 4 x_{t} y )}{4(-1 + x_{t})(x_{t} - y) y} + \frac{3 x_{t}^{3}(x_{t} - 2 y + x_{t} y) \log{x_{t}}}{4(-1 + x_{t})^{2}(x_{t} - y)^{2} y} \nonumber \\
& - & \frac{3 x_{t}^2 y \log{y}}{4(x_{t} - y)^{2} (-1 + y)} \\
\nonumber\\
P_{4}(x_{c}, x_{t}, y)\;\,\,\;\;\;\; & = & \frac{3 x_{c} x_{t} y \log{y}}{4(x_{t} - y) (-1 + y) (y - x_{c})} + \frac{(-4 x_{t} + x_{t}^{2} + 4 y - 4 x_{t} y) x_{c} x_{t} \log{x_{t}}}{4(-1 + x_{t}) (x_{t} - y)(x_{t} - x_{c}) y} \nonumber \\
& - & \frac{(-4 y + 4 x_{c} + 4 y x_{c} - x_{c}^{2}) x_{c} x_{t}
  \log{x_{c}}}{4(-1 + x_{c}) (y - x_{c})(x_{t} - x_{c}) y} \\
\nonumber\\
P_5(x_t, x_T, y)\,\,\;\;\;\;&=&
   - \frac{x_t(-3x_t^4 + 4x_t^2 x_T - 2 x_t^3 x_T + x_t^4 x_T + 6
     x_t^3y - 3x_t^4y - 8 x_t x_T y) \log x_t}
     {4{( -1 + x_t ) }^2
     ( x_t - x_T ) {( x_t - y ) }^2 y}\nonumber\\  
&-& \frac{x_t(+7x_t^2 x_T y - 2x_t^3x_T y + 4 x_T y^2 - 8x_t x_T y^2 +
  4 x_t^2 x_T y^2 ) \log x_t}
     {4{( -1 + x_t ) }^2
     ( x_t - x_T ) {( x_t - y ) }^2 y}\nonumber\\
   &+& \frac{x_t x_T( -4x_T + x_T^2 + 4 y - 4 x_T y
       ) \log x_T}{4 ( x_t - x_T ) 
     ( -1 + x_T ) ( x_T - y ) y} + 
  \frac{3 x_t ( x_t - x_T  ) y^2 \log y}
   {4 { ( x_t - y ) }^2( x_T - y ) 
     ( -1 + y ) }\nonumber\\
&+&\frac{x_t( -4 x_t + x_t^2 + 4y - 4x_ty ) }
   {4( -1 + x_t) ( x_t - y ) y}\\
\nonumber\\
P_6(x_c, x_t, x_T, y)&=&
 -\frac{x_c x_t( x_t^2 + 4 y - 
       4 x_t( 1 + y)  ) \log x_t}
     {4 ( x_c - x_t ) ( -1 + x_t) ( x_t - y ) y} 
+ \frac{x_c x_T( x_T^2 + 4y - 
       4 x_T ( 1 + y )  )\log x_T}
     {4( x_c - x_T ) ( -1 + x_T ) 
     ( x_T - y )y}\nonumber\\
&+&\frac{x_c^2(x_t - x_T) ( x_c^2 + 4y - 4 x_c ( 1 + y ) 
       ) \log x_c}{4 ( -1 + x_c ) ( x_c - x_t ) ( x_c - x_T) ( x_c - y ) y}\nonumber\\
& + &
  \frac{3x_c( x_t - x_T ) y^2\log y}
   {4( x_c - y ) ( -1 + y ) 
     ( -x_t + y ) ( -x_T + y ) }\\
\nonumber\\
P_{7}(x_{t}, z)\;\;\;\;\;\;\;\;\;\;\; & = & \frac{x_{t}^{2}(-4 + x_{t})}{4(-1 + x_{t})(x_{t} - z)} - \frac{x_{t}^{2}(-3 x_{t}^{2} + 4 z - 2 x_{t} z + x_{t}^{2} z) \log{x_{t}}}{4(-1 + x_{t})^{2}(x_{t} - z)^{2}} \nonumber \\
& + & \frac{x_{t}^{2}(-4 z + z^{2}) \log{z}}{4(x_{t} - z)^{2} (-1 + z)}\\
P_{TT}(x_{t}, x_{T})\;\;\;\;\,\,\,\, & = & \,\, \frac{x_{T}}{4} + \frac{-9 + 16 x_{t} - 14 x_{t}^{2} + x_{t}^{3}}{4(-1 + x_{t})^{2}} - \frac{6}{4(-1 + x_{T})^{2}} - \frac{3 (-5 + 3 x_{t})}{4 (-1 + x_{t})(-1 + x_{T})} \nonumber \\
& - & \,\, \frac{x_{t} (-3 x_{t}^{3} - 4 x_{T} + 12  x_{t} x_{T} - 6 x_{t}^{2} x_{T} + x_{t}^{3} x_{T}) \log{x_{t}}}{2 (-1 + x_{t})^{3} (x_{t} - x_{T})} \nonumber \\
& + & \,\, \frac{x_{T} (- 4 x_{t} + 12 x_{t} x_{T} - 6 x_{t} x_{T}^{2} - 3 x_{T}^{3} + x_{t} x_{T}^{3}) \log{x_{T}}}{2(x_{t} - x_{T})(-1 + x_{T})^{3}}
\end{eqnarray}

\end{appendix}

%
%
%



%
%
%
\end{document}